\documentclass[11pt,letterpaper]{article}
\usepackage[normalem]{ulem}

\usepackage[margin=1in]{geometry}
\usepackage[T1]{fontenc}
\usepackage{microtype}
\usepackage{parskip}
\usepackage{setspace}
\usepackage[sc]{mathpazo}        
\usepackage[scale=0.95]{tgheros} 
\usepackage{sectsty}

\usepackage{graphicx}
\usepackage[table]{xcolor}
\usepackage{booktabs}
\usepackage{tabularx}
\usepackage{longtable}
\usepackage{array}
\usepackage[export]{adjustbox}
\usepackage{makecell}
\usepackage{float}

\usepackage{caption}
\usepackage[backend=biber,style=numeric,sorting=none,doi=true,url=true]{biblatex}
\usepackage{url}
\usepackage{hyperref}
\hypersetup{
  colorlinks=true,
  linkcolor=blue!50!black,
  urlcolor=blue!50!black,
  citecolor=blue!50!black,
  pdfauthor={Shams El-Adawy et al.},
  pdftitle={Profiles of Roles in the Quantum Industry}
}

\usepackage{titletoc}
\titlecontents{section}
  [0em]{\vspace{2pt}}{\contentslabel{1.5em}}{}{\titlerule*[.5pc]{.}\contentspage}
\titlecontents{subsection}
  [1.5em]{\vspace{1pt}}{\contentslabel{1.8em}}{}{\titlerule*[.5pc]{.}\contentspage}
\titlecontents{subsubsection}
  [3.0em]{\vspace{1pt}}{\contentslabel{2.3em}}{}{\titlerule*[.5pc]{.}\contentspage}

\usepackage{enumitem}
\setlist{nosep,leftmargin=1.2em}

\usepackage[most]{tcolorbox}
\newtcolorbox{questionbox}{
  enhanced, breakable,
  colback=gray!8, colframe=gray!50,
  borderline west={3pt}{0pt}{gray!60},
  boxrule=0.5pt, arc=2mm,
  left=8pt, right=8pt, top=8pt, bottom=8pt,
  fonttitle=\sffamily\bfseries, coltitle=black,
  title={\sffamily\bfseries Key Question}
}

\AtBeginEnvironment{quote}{\small\itshape}

\usepackage{pagecolor}

\definecolor{brandnavy}{HTML}{243B7A}
\definecolor{hardwaredark}{HTML}{4E6E5D}
\definecolor{hardwarelight}{HTML}{E8F0EC}
\definecolor{softwaredark}{HTML}{F58F29}
\definecolor{bridgingdark}{HTML}{AD2E24}
\definecolor{publicdark}{HTML}{0576B3}
\providecommand{\seriesfontsize}{\normalsize}
\providecommand{\seriesitem}[3]{}
\providecommand{\seriesitemcurrent}[3]{}
\providecommand{\seriesstripe}[5]{}

\renewcommand{\seriesfontsize}{\large} 
\newcommand{\serieslinesep}{12pt}       

\renewcommand{\seriesitem}[3]{%
  \noindent
  \begin{minipage}[t]{0.72\linewidth}
    {\sffamily\seriesfontsize\color{white} Report #1:\enspace #2}%
  \end{minipage}%
  \begin{minipage}[t]{0.28\linewidth}\raggedleft
    {\sffamily\seriesfontsize\color{white!85} #3}%
  \end{minipage}\par\vspace{\serieslinesep}%
}

\renewcommand{\seriesitemcurrent}[3]{\seriesitem{#1}{#2}{#3}}

\renewcommand{\seriesstripe}[5]{%
  \begin{tcolorbox}[enhanced, breakable,
    colback=brandnavy, colframe=brandnavy,
    boxrule=0pt, sharp corners,
    left=0pt, right=0pt, top=4pt, bottom=10pt,
    borderline south = {1pt}{0pt}{white!80}
  ]
    \noindent
    \begin{minipage}[t]{0.62\linewidth}
      {\sffamily\small\color{white!85}\MakeUppercase{#1}}%
    \end{minipage}%
    \begin{minipage}[t]{0.38\linewidth}\raggedleft
      {\sffamily\small\color{white!85} Publication date}%
    \end{minipage}

    \par\vspace{3pt}{\color{white!70}\rule{\linewidth}{0.4pt}}\vspace{4pt}

    \seriesitemcurrent{#2}{#4}{#3}

    {\sffamily\seriesfontsize\color{white!85} }\par\vspace{2pt}
    #5
  \end{tcolorbox}%
}

\sectionfont{\sffamily\bfseries\color{brandnavy}}
\subsectionfont{\sffamily\bfseries\large\color{brandnavy}}
\subsubsectionfont{\sffamily\bfseries\normalsize\color{brandnavy}}

\usepackage{fancyhdr}
\renewcommand{\headrulewidth}{0.2pt}
\makeatletter
\renewcommand{\headrule}{%
  \vspace{-3pt}\hbox to\headwidth{\color{brandnavy}\leaders\hrule height \headrulewidth\hfill}}
\makeatother

\usepackage{fontawesome5}

\newtcolorbox{profilecard}[2][]{%
  enhanced, breakable, sharp corners,
  colback=hardwarelight, colframe=hardwaredark!60,
  left=10pt, right=10pt, top=8pt, bottom=8pt,
  boxrule=0.4pt, arc=2mm,
  borderline west={3pt}{0pt}{hardwaredark},
  fonttitle=\sffamily\bfseries\color{hardwaredark},
  title={\sffamily\bfseries\large #2}, #1
}

\newcommand{\ProfileSectionHeader}[3][book]{%
  \vspace{1em}
  \begin{center}
    {\sffamily\bfseries\large\textcolor{#2}{\faIcon{#1}\enspace #3}}\\[-2pt]
    {\color{gray!40}\rule{0.55\linewidth}{0.6pt}}\\[4pt]
  \end{center}
}

\newcommand{\ProfileBlockSpacing}{\vspace{0.8em}}
\newcommand{\fieldlabel}[1]{\textbf{#1}}

\newcommand{\HardwareProfileBlock}[6]{%
\begin{tcolorbox}[
  enhanced, breakable, sharp corners,
  colback=hardwarelight,
  colframe=hardwaredark!60,
  boxrule=0.4pt,
  arc=2mm,
  borderline west={3pt}{0pt}{hardwaredark},
  left=10pt, right=10pt, top=8pt, bottom=8pt,
  title={\centering\sffamily\large\textcolor{white}{#1}},
  fonttitle=\sffamily\large,
  coltitle=white,
]
\begin{center}
  {\scriptsize\textcolor{gray!60}{\itshape Findings in this profile are not presented in order of importance, nor does any individual position in the data include all listed elements.}}
\end{center}
\vspace{4pt}
{\small
\begin{itemize}[label={}, leftmargin=1.5em, itemsep=1pt, topsep=1pt]
  \item \textbf{Individual positions:} #2
  \item \textbf{Company types:} #3
\end{itemize}
}

\ProfileSectionHeader[building]{hardwaredark}{Occupation-Specific Information}
#4

\ProfileSectionHeader[brain]{hardwaredark}{Worker Requirements: Knowledge, Skills, Abilities (KSAs)}
#5

\ProfileSectionHeader[graduation-cap]{hardwaredark}{Experience Requirements}
#6
\end{tcolorbox}
}

\newcommand{\logorowscale}{0.90}
\newcommand{\logospace}{1.2cm}

\title{%
  \vspace{-1.2cm}%
  \resizebox{\logorowscale\textwidth}{!}{%
    \makebox[\textwidth]{%
      \adjincludegraphics[valign=c,height=0.66cm]{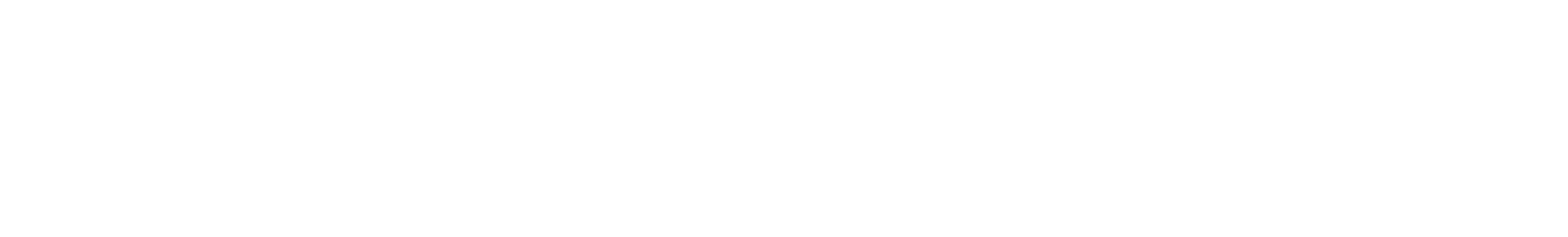}%
      \hspace{\logospace}%
      \adjincludegraphics[valign=c,height=0.60cm]{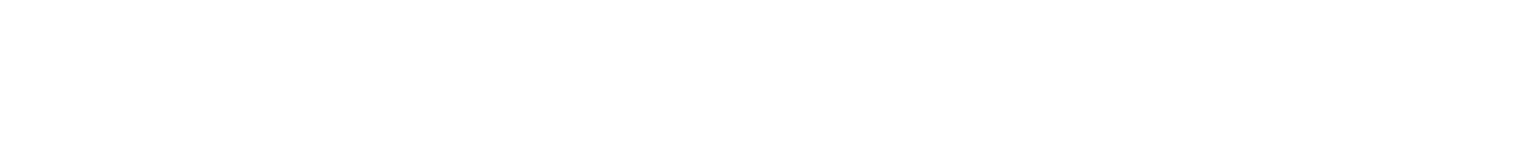}%
    }%
  }\\[1.2cm]
  {\sffamily\bfseries\Huge Profiles of Roles in the Quantum Industry}\\[0.6cm]
  {\sffamily\large Report 3 — January 2026}
}

\author{%
  \parbox{\textwidth}{\centering
    \begingroup
    \setlength{\tabcolsep}{18pt}%
    \begin{tabular*}{0.9\textwidth}{@{\extracolsep{\fill}} c c}
      \makecell{\textbf{Shams El-Adawy}\\ \textbf{Heather J. Lewandowski}\\[2pt]\small University of Colorado Boulder} &
      \makecell{\textbf{A.\,R. Pi\~{n}a}\\ \textbf{Benjamin M.\,Zwickl}\\[2pt]\small Rochester Institute of Technology}
    \end{tabular*}
    \endgroup
  }%
}
\date{}

\begin{document}

\begingroup
  \pagecolor{brandnavy}\color{white}
  \hypersetup{linkcolor=white!85, urlcolor=white!85, citecolor=white!85}

  \maketitle

  \begin{tcolorbox}[enhanced, breakable,
    colback=brandnavy, colframe=brandnavy,
    boxrule=0pt, sharp corners,
    left=0pt, right=0pt, top=4pt, bottom=10pt,
    borderline south = {1pt}{0pt}{white!80}
  ]
    \noindent
    \begin{minipage}[t]{0.62\linewidth}
      {\sffamily\small\color{white!85}\MakeUppercase{Quantum Workforce Report Series}}%
    \end{minipage}%
    \begin{minipage}[t]{0.38\linewidth}\raggedleft
      {\sffamily\small\color{white!85} Publication date}%
    \end{minipage}

    \par\vspace{3pt}{\color{white!70}\rule{\linewidth}{0.4pt}}\vspace{6pt}

    {\sffamily\large\color{white!85} }\par\vspace{4pt}

    \seriesitem{1}{\href{https://arxiv.org/abs/2510.12936}{Experimental Skills for Non-PhD Roles in the Quantum Industry}}{October 2025}\seriesitem{2}{\href{https://arxiv.org/abs/2511.11820}{Categorization of Roles in the Quantum Industry}}{November 2025}
    \seriesitem{3}{Profiles of Roles in the Quantum Industry}{January 2026}%
  \end{tcolorbox}

  \thispagestyle{empty}
\endgroup

\clearpage
\nopagecolor
\color{black}
\hypersetup{linkcolor=blue!50!black, urlcolor=blue!50!black, citecolor=blue!50!black}

\tableofcontents
\thispagestyle{empty}
\clearpage

\section*{Executive Summary}
\addcontentsline{toc}{section}{Executive Summary}
This report builds upon the \textit{Categorization of Roles in the Quantum Industry} \cite{Pina2025Categorization} report by providing detailed profiles for 29 distinct roles across the quantum workforce. While the earlier report established a framework of four major role categories (hardware, software, bridging, and public-facing and business) and their subcategories, the current report expands on this structural framework by characterizing what professionals in each role actually do, particularly by identifying the tasks, knowledge, skills, abilities (KSAs), and experience typically required for each role.
Each role profile follows a standardized structure guided by the Occupational Information Network (O*NET) framework \cite{onet_content_model}, which includes:
\begin{itemize}
    \item Profile header: role title, title of individual positions contributing to the profile, and associated company types
    \item Occupation-specific information: description of role and key tasks
    \item Worker requirements: technical and general knowledge, skills, and abilities
    \item Experience requirements: education, prior experience, and on-the-job training information
\end{itemize}
By presenting a fine-grained view of day-to-day work and qualification expectations, this report serves as a practical resource for educators, students, industry professionals, and policymakers aiming to understand, educate, and support the evolving quantum workforce.

\vspace{0.5cm} \textbf{Recommendations:} \begin{enumerate}

   \item \textbf{Align education with role-specific tasks and KSAs}:  Educators can use the detailed role profiles to inform course and program design across the quantum education ecosystem. By mapping curriculum content to specific tasks and KSAs identified, educators can ensure that students develop skills directly aligned with workforce needs.  
   \item \textbf{Use the profiles to inform workforce development strategies}: Policymakers can use these profiles for identifying areas where investment is needed.  
   \item \textbf{Support career exploration and planning}: Students can use the role profiles to better understand career pathways within the quantum industry. These profiles provide insights into the educational backgrounds, training, and skills typically associated with various roles, which can help students make informed decisions about their academic and professional development. 
\end{enumerate}

\clearpage

\section*{How to Use This Report}
\addcontentsline{toc}{section}{How to Use This Report}

This report is designed to be used as a reference rather than read linearly from beginning to end. It may be useful to read it in parallel with \textit{Report 2: Categorization of Roles in the Quantum Industry} \cite{Pina2025Categorization}, which provides the structural framework for these profiles.
Readers will generally find it most useful to navigate directly to the roles that are relevant to their interests.
\begin{itemize}
  \item \textbf{Start with the first few pages for context:} 
  The motivation, data collection, and analysis sections provide an overview of the work, data, and its limitations.
  \item \textbf{Use the table of contents as your main entry point:} All items in the Table of Contents are clickable links. Selecting the title of a section, subgroup, or role profile allows you to jump directly to that part of the report.
\end{itemize}

\clearpage

\section*{Motivation}
\addcontentsline{toc}{section}{Motivation}

As the field of Quantum Information Science and Engineering (QISE) continues to expand, there is an increasing need to understand what knowledge, skills, and abilities (KSAs) quantum professionals actually make use of in their day-to-day work \cite{el2025insights}. This understanding is needed to design education that prepares individuals to enter  the quantum workforce. 

Role profiles provide one way to meet this need. By bringing together information from multiple individual positions, the profiles provide a synthesized view of a role that highlights the range of tasks, skills, knowledge, and experiences needed. Unlike job descriptions, which describe the specific duties of a single position within one organization, role profiles capture the range of similar work within a domain area across the quantum workforce. This approach offers a more comprehensive picture of the quantum workforce and can inform education and workforce development efforts. 

Recent studies \cite{goorney2025quantum,devendrababu2025mapping} have provided valuable quantitative insights into workforce needs by systematically analyzing job postings. Those findings illustrate large scale trends, but primarily reflect the perspective of employers as expressed in job advertisements. In contrast, this report offers a complementary perspective by focusing on the lived experiences of professionals actively working in the quantum industry. Whereas job postings capture how employers describe a position they seek to fill, role profiles in this report capture how multiple professionals describe the work they actually perform or supervise. This complementary approach provides depth and context that may not always be inferred from job advertisements alone. 

This report follows and expands the report on \textit{Categorization of Roles in the Quantum Industry} \cite{Pina2025Categorization}, which characterized the organization of roles in the quantum industry. In particular, that report provided the structural framework of four major role groups: hardware, software, bridging, and public-facing and business. The current report builds upon that categorization by developing comprehensive profiles for each role in each group. This offers a finer-grained view of the day-to-day tasks, required technical and general knowledge, skills, and abilities (KSAs), and experience associated with 29 roles across the quantum workforce.
\begin{tcolorbox}[
  enhanced, breakable,
  colback=gray!10,
  colframe=gray!50,
  boxrule=0.4pt,
  arc=2mm,
  left=10pt, right=10pt, top=8pt, bottom=8pt,
  borderline west={3pt}{0pt}{gray!60},
  fonttitle=\sffamily\bfseries,
  title={\sffamily\bfseries Goal of this report}
]

Describe what individuals in each role within a group actually do by identifying the day-to-day tasks, the knowledge, skills, and abilities (KSAs), and the experience typically required for each role

\end{tcolorbox}

\section*{Data Collection and Analysis}
\addcontentsline{toc}{section}{Data Collection and Analysis}

The data collection mirrors the dataset outlined in \textit{Report 2: Categorization of Roles in the Quantum Industry} \cite{Pina2025Categorization}. We restate that information, and then present the analysis relevant for profile development. 

\begin{figure}[H]
    \centering
\includegraphics[width=0.9\linewidth]{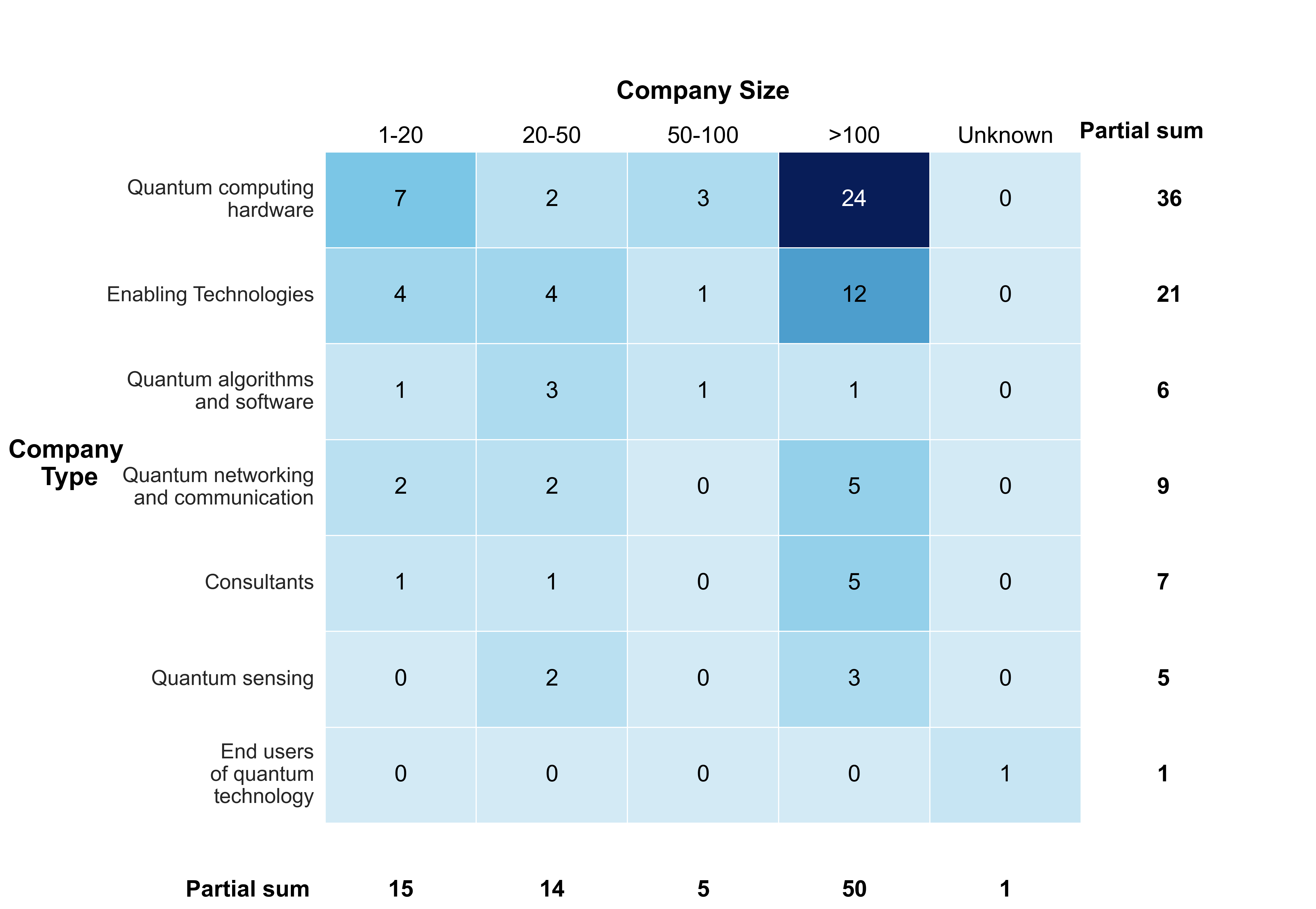}
    \caption{Distribution of interviewed companies by type and size.  We did 42 interviews with professionals at 23 distinct companies. We had a few interviewees from the same company and most companies self-reported participation in more than one type of activity. We adopted definitions consistent with prior literature \cite{fox2020preparing, ElAdawy2025} for company types. Company size corresponds to the range of employees working in quantum-related technologies in these companies.}
    \label{fig:company_distribution}
\end{figure}

The data collected for this work were semi-structured interviews with professionals currently employed in the QISE industry. 
These individuals worked at companies spanning most sectors of the industry, however, some are more represented than others (see Figure \ref{fig:company_distribution}). 
The interviews were conducted with two protocols, both of which included questions about the tasks, knowledge, skills, abilities, and educational requirements associated with one or more quantum-related positions at the participants' company. 
One protocol was intended for managers who could speak to a variety of positions within their company and the other was intended for employees  and focused on their specific position. 
Although the interviews did include other topics related to the QISE industry, these were the most pertinent questions in creating the profiles discussed here.

\clearpage
\textbf{Profile development process:}
\begin{itemize}
    \item We identified and extracted relevant information from the interview transcripts corresponding to specific components of the ONET framework \cite{onet_content_model}. 
    \item Using the data characterized in the framework, we drafted each part of the profile in the framework, including occupation-specific information, worker requirements, and experience requirements. 
    \item Each profile was developed using data from as many individual positions as we had available as described in our role categorization process. We note that no single position encompasses all the characteristics described in one role, unless the role was developed based on a single individual position. For most profiles, the profile represents a combination of information from several individual positions that share common tasks.
    \item The text in the profiles was written to closely reflect the language used by participants during interviews to increase accuracy and minimize misinterpretations.
    \item Once profile drafts were completed, we reviewed each profile for consistency in language and structure. 
\end{itemize}

\vspace{0.5cm}

\begin{tcolorbox}[
  enhanced, breakable,
  colback=gray!5,
  colframe=gray!50,
  boxrule=0.4pt,
  arc=2mm,
  borderline west={3pt}{0pt}{gray!60},
  left=10pt, right=10pt, top=8pt, bottom=8pt,
  title={\centering\sffamily\large\textcolor{black}{\emph{Template of a Profile: Role Name}}}
]
\small

\emph{Individual positions:} The titles of the individual positions that informed the profile

\emph{Company types:} The types of companies represented; for the quantum computing hardware companies, we indicate, where possible, the type platform they use: trapped-ion, neutral-atom, or superconducting.

\ProfileBlockSpacing

\ProfileSectionHeader[building]{brandnavy}{Occupation-Specific Information}
\emph{Description:} Description of what people in the role do

\emph{Tasks:} Key day-to-day tasks

\ProfileBlockSpacing

\ProfileSectionHeader[brain]{brandnavy}{Worker Requirements: Knowledge, Skills, Abilities (KSAs)}
\emph{Knowledge of:} Knowledge needed for this role listed in alphabetical order

\emph{Occupation-specific skills and abilities:} Specific technical skills and abilities used in the role listed in alphabetical order

\emph{General skills and abilities:} General skills and abilities used in this role listed in alphabetical order

\ProfileBlockSpacing

\ProfileSectionHeader[graduation-cap]{brandnavy}{Experience Requirements}
\emph{Education:} Information provided by interviewees about common degrees and disciplines for this role

\emph{On-the-job training:} Information provided by interviewees about on-the-job training for this role

\emph{Prior experience:} 
Information provided by interviewees about necessary or preferred prior experience for this role

\end{tcolorbox}

\smallskip
{\normalsize\textit{%
  \textcolor{brandnavy}{Note:} In the profile cards throughout this report, 
  the background color and section headers indicate the role category: 
  \textcolor{hardwaredark}{\textbf{Hardware roles}}, 
  \textcolor{softwaredark}{\textbf{Software roles}}, 
  \textcolor{bridgingdark}{\textbf{Bridging roles}}, and 
  \textcolor{publicdark}{\textbf{Public-facing and business roles}}.}}

\vspace{0.5cm} \textbf{Limitations:}

There are some limitations in the profiles presented, which we hope to address in future work by conducting additional interviews to refine these profiles: 
\begin{itemize}
    \item \textit{Scope of data}: Some profiles are based on a small number of interviews, and may not capture the full breath of responsibilities or experiences associated with each role. 
    \item  \textit{Variability in detail}: There are varying levels of specificity in the KSAs, which reflects differences in how participants discussed their work. Some profiles include highly specific information, while others include more general descriptions. This variation may reflect the limitation of the data collected, rather than the requirements for the actual positions within a role profile. 
    \item \textit{Challenges in  characterizing the depth of knowledge and skills}: There is no universal standard for defining what constitutes basic versus advanced KSAs. Participants’ descriptions often reflected a spectrum of topic areas and skills rather than fixed expertise levels, which makes it challenging to consistently establish distinctions in knowledge depth without considering the corresponding tasks. For clarity and consistency, we present tasks and KSAs separately in the profiles, even though they are best interpreted together.   
\end{itemize}

\newcommand{\ProfileHardwareFifteen}{%
\HardwareProfileBlock
{Profile H1.1: Senior Scientists}
{Principal Investigator, Senior Advanced Physicist, Senior Engineer Optical System Integration}
{Quantum computing hardware (platforms: trapped-ion, neutral-atom), Quantum algorithms and software, Quantum sensing, Quantum networking and communication, Enabling technologies}
{%
  \fieldlabel{Description:} Senior scientists work in areas closely connected to fundamental research. They oversee technical projects and manage people.

  \vspace{0.5em}
  \fieldlabel{Tasks:} On the job, they would…
  \begin{itemize}
    \item Coordinate teams with different areas of expertise
    \item Drive innovation
    \item Handle hiring and training of engineering staff
    \item Implement strategies proposed by theory teams
    \item Lead programs
    \item Oversee integration of subsystems into a final product
    \item Write proposals
  \end{itemize}
}
{%
  \fieldlabel{Knowledge of:}
\begin{itemize}
    \item AMO (Atomic, Molecular, and Optical) physics
    \item Ion trapping
    \item Laser optics
    \item Quantum algorithm development
    \item Quantum computing theory 
    \item Quantum error correction and decoherence
    \item Quantum gates
    \item Quantum hardware development
    \item Quantum imaging techniques 
    \item Quantum system design and control
    \item Quantum technology implementation
    \item Vacuum systems
\end{itemize}

  \fieldlabel{Occupation-specific skills and abilities:}
\begin{itemize}
    \item Able to collaborate and communicate effectively across disciplines
    \item Able to integrate hardware components for engineering systems
    \item Able to manage large-scale projects and coordinate teams
    \item Able to solve problems independently and as part of a team
    \item Able to use programming languages for data collection, analysis, and system control
\end{itemize}

  \fieldlabel{General skills and abilities:}
  \begin{itemize}
    \item Critical thinking and troubleshooting
  \end{itemize}
}
{%
  \fieldlabel{Education:} 
  \begin{itemize}
  \item Degree: PhD
  \item Disciplines: Electrical Engineering or Physics
  \end{itemize}
  \fieldlabel{On-the-job training:} 
  \begin{itemize}
  \item None beyond orientation
  \end{itemize}

  \fieldlabel{Prior experience:}
  \begin{itemize}
    \item Often 10+ years of industry experience is required
    \item PhD and postdoctoral experience are minimum requirements
  \end{itemize}
}
}

\clearpage
\newcommand{\ProfileHardwareSixteen}{%
\begin{tcolorbox}[
  enhanced, breakable, sharp corners,
  colback=hardwarelight,
  colframe=hardwaredark!60,
  boxrule=0.4pt,
  arc=2mm,
  borderline west={3pt}{0pt}{hardwaredark},
  left=10pt, right=10pt, top=8pt, bottom=8pt,
  title={\centering\sffamily\large\textcolor{white}{Profile H1.2: Engineering Managers}},
  fonttitle=\sffamily\large,
  coltitle=white,
]

\begin{center}
  {\scriptsize\textcolor{gray!}{\itshape Findings in this profile are not presented in order of importance and will be refined as additional interviews are conducted.}}
\end{center}

\vspace{4pt}
{\small
\begin{itemize}[label={}, leftmargin=1.5em, itemsep=1pt, topsep=1pt]
  \item \textbf{Individual positions:} Quantum R\&D Engineer
  \item \textbf{Company types:} Quantum computing hardware (platform: neutral-atom), Quantum algorithms and software, Quantum sensing, Quantum networking and communication, Enabling technologies
\end{itemize}
}

\ProfileSectionHeader[building]{hardwaredark}{Occupation-Specific Information}
\fieldlabel{Description:} Engineering managers oversee teams of engineers building hardware systems.

\vspace{0.5em}
\fieldlabel{Tasks:} On the job, they would...
\begin{itemize}
  \item Build proprietary equipment
  \item Generate requirements for engineers
  \item Integrate and test multiple systems
  \item Perform demonstrations and showcase applications of products
  \item Train others to use and service proprietary equipment
\end{itemize}

\ProfileSectionHeader[brain]{hardwaredark}{Worker Requirements: Knowledge, Skills, Abilities (KSAs)}
\fieldlabel{Knowledge of:}
\begin{itemize}
  \item Commonly used experimental techniques
  \item Non-quantum electronics
  \item Optics
  \item Quantum mechanics
  \item Quantum system design and control
  \item RF (Radio Frequency) electronics
  \item Systems engineering and project integration
  \item Thermodynamics
\end{itemize}

\fieldlabel{Occupation-specific skills and abilities:}
\begin{itemize}
  \item Able to communicate effectively across teams
  \item Able to do some programming generally, without specifying whether it is for data processing, data representation, or other purposes
  \item Able to document all practices
  \item Able to implement various experimental techniques common in the field in different settings
  \item Able to perform rigorous device testing 
  \item Able to read and interpret scientific papers
  \item Able to translate quantum plans  into requirements for the various subsystems that will support the experiment
\end{itemize}

\fieldlabel{General skills and abilities:}
\begin{itemize}
  \item Program management
\end{itemize}

\ProfileSectionHeader[graduation-cap]{hardwaredark}{Experience Requirements}
\fieldlabel{Education:} 
\begin{itemize}
    \item Degree: Bachelor 
    \item Disciplines: Engineering or Physics
\end{itemize}

\fieldlabel{On-the-job training:}
\begin{itemize}
  \item Certifications from professional societies
  \item Conference attendance
\end{itemize}

\fieldlabel{Prior experience:}
\begin{itemize}
  \item Four years of relevant industry experience can substitute for education requirements
   \item No prior experience needed if education requirements are met
\end{itemize}

\end{tcolorbox}
}

\clearpage
\newcommand{\ProfileHardwareSeventeen}{%
\HardwareProfileBlock
{Profile H2.1: Commercialization Leads}
{Staff Scientist, Quantum Scientist, Quantum Engineer}
{Quantum computing hardware (platforms: trapped-ion, neutral-atom), Quantum algorithms and software, Enabling technologies}
{%
  \fieldlabel{Description:} Commercialization leads guide a team and drive product development, commercialization, and applications. They coordinate and contribute to the design, construction, and operation of quantum technologies.

  \vspace{0.5em}
  \fieldlabel{Tasks:} On the job, they would...
  \begin{itemize}
    \item Coordinate multidisciplinary teams
    \item Design and develop new products
    \item Model components using CAD (Computer-Aided Design) software
    \item Implement and test new features
    \item Run client jobs on quantum computers
    \item Write software to control experiments
    \item Write reports and publications
  \end{itemize}
}
{%
  \fieldlabel{Knowledge of:}
\begin{itemize}
    \item Non-quantum electronics
    \item Optical physics
    \item Quantum algorithms
    \item Quantum computing
    \item Quantum information science 
    \item Quantum mechanics
\end{itemize}

  \fieldlabel{Occupation-specific skills and abilities:}
  \begin{itemize}
    \item Able to do some programming generally, without specifying whether it is for data processing, data representation, or other purposes
    \item Able to use CAD software
  \end{itemize}

  \fieldlabel{General skills and abilities:}
\begin{itemize}
    \item Communication, data analysis, problem solving, project management, and troubleshooting
\end{itemize}

}
{%
  \fieldlabel{Education:} 
  \begin{itemize}
  \item Degree: PhD
  \item Disciplines: Chemistry or Physics
  \end{itemize}
  
  \fieldlabel{On-the-job training:}
  \begin{itemize}
    \item One-on-one mentorship for technical skills
    \item Group trainings on managing teams of people
    \item Group trainings on managing projects effectively
  \end{itemize}

  \fieldlabel{Prior experience:}
  \begin{itemize}
    \item About 5 years postdoc or industry experience is preferred
    \item Often can be hired directly after PhD or postdoc
  \end{itemize}
}
}

\clearpage
\newcommand{\ProfileHardwareEighteen}{%
\begin{tcolorbox}[
  enhanced, breakable, sharp corners,
  colback=hardwarelight,
  colframe=hardwaredark!60,
  boxrule=0.4pt,
  arc=2mm,
  borderline west={3pt}{0pt}{hardwaredark},
  left=10pt, right=10pt, top=8pt, bottom=8pt,
  title={\centering\sffamily\large\textcolor{white}{Profile H2.2: Systems Engineers}},
  fonttitle=\sffamily\large,
  coltitle=white,
]

\begin{center}
  {\scriptsize\textcolor{gray!}{\itshape Findings in this profile are not presented in order of importance and will be refined as additional interviews are conducted.}}
\end{center}

\vspace{4pt}
{\small
\begin{itemize}[label={}, leftmargin=1.5em, itemsep=1pt, topsep=1pt]
  \item \textbf{Individual positions:} System Integration Engineer
  \item \textbf{Company types:} Quantum computing hardware (platform: trapped-ion), Quantum algorithms and software
\end{itemize}
}

\ProfileSectionHeader[building]{hardwaredark}{Occupation-Specific Information}
\fieldlabel{Description:} Systems engineers perform traditional systems engineering in a quantum setting. For example, when building a quantum computer, they must understand all of the systems and sub-systems, how they interact with one another, how they interface with relevant software, and how they are impacted by environmental factors.

\vspace{0.5em}
\fieldlabel{Tasks:} On the job, they would…
\begin{itemize}
    \item Document all of the subsystems, hardware-software interactions, and environmental interactions
  \item Manage the design and construction of a quantum computer
  \item Manage testing and data analysis procedures
\end{itemize}

\ProfileSectionHeader[brain]{hardwaredark}{Worker Requirements: Knowledge, Skills, Abilities (KSAs)}
\fieldlabel{Knowledge of:}
\begin{itemize}
  \item AMO physics 
  \item Non-quantum electronics
  \item Statistics 
\end{itemize}

\fieldlabel{Occupation-specific skills and abilities:}
\begin{itemize}
  \item Able to apply experimental skills in quantum computing development context
  \item Able to troubleshoot hardware and software
  \item Able to work with ambiguity and adapt to changing circumstances
  \item Able to work with optics 
\end{itemize}

\fieldlabel{General skills and abilities:}
\begin{itemize}
  \item Critical thinking, data analysis, and problem solving
\end{itemize}

\ProfileSectionHeader[graduation-cap]{hardwaredark}{Experience Requirements}
\fieldlabel{Education:} 
\begin{itemize}
\item Degree: PhD
\item Discipline: Physics 
\end{itemize}

\fieldlabel{On-the-job training:} 
\begin{itemize}
\item One-on-one training as needed
\end{itemize}

\fieldlabel{Prior experience:} 
\begin{itemize}
\item Experience in engineering and business management are helpful but not required
\end{itemize} 

\end{tcolorbox}
}

\clearpage
\newcommand{\ProfileHardwareNineteen}{%
\HardwareProfileBlock
{Profile H3.1: Experimental Scientists}
{Quantum Optics Engineer, Quantum Hardware Scientist, Research Scientist, Senior Advanced Physicist, Quantum Experimental Algorithms Researcher, Senior Photonics Experimenter}
{Quantum computing hardware (platforms: trapped-ion, superconducting), Quantum algorithms and software, Quantum sensing, Quantum networking and communication, Enabling technologies}
{%
  \fieldlabel{Description:} Experimental scientists plan and run the experiments (e.g., doing quantum optics experiments, device testing), perform analysis, and report results either internally (e.g., report) or externally (e.g., publications). 

  \vspace{0.5em}
  \fieldlabel{Tasks:} On the job, they would...
  \begin{itemize}
    \item Characterize hardware performance by taking and analyzing data
    \item Communicate results and plans within the organization
    \item Contribute to relocation of large scale quantum experiments
    \item Perform and plan experiments
    \item Review scientific literature
    \item Write reports and publications
  \end{itemize}
}
{%
  \fieldlabel{Knowledge of:}
\begin{itemize}

    \item Decoherence
    \item Device physics
    \item Error mitigation techniques
    \item How quantum networks scale in practice
    \item Ion trapping
    \item Non-quantum electronics
    \item Photon counting and device physics
    \item Probability theory
    \item Qubit hardware
    \item Quantum algorithms
    \item Quantum design and control
    \item Quantum imaging techniques
    \item Quantum networking and optics
    \item Quantum technology implementation
\end{itemize}

  \fieldlabel{Occupation-specific skills and abilities:}
  \begin{itemize}
    \item Able to align optical elements
    \item Able to characterize and calibrate lasers
    \item Able to communicate effectively about technical topics
    \item Able to do some programming generally, without specifying whether it is for data processing, data representation, or other purposes
    \item Able to network multiple computing platforms together
    \item Able to perform quantum optics experiments
    \item Able to troubleshoot hardware and software
    \item Able to work with vacuum systems
\end{itemize}

  \fieldlabel{General skills and abilities:}
  \begin{itemize}
    \item Adaptability, collaboration, critical thinking, leadership, problem solving, and project management
  \end{itemize}
}
{%
  \fieldlabel{Education:} 
  \begin{itemize}
  \item Degrees: Bachelor, Master, or PhD 
  \item Disciplines: Computer Science, Engineering (mechanical or thermal engineering), or Physics
  \end{itemize}

  \fieldlabel{On-the-job training:}
  \begin{itemize}
    \item Colloquia and journal clubs
    \item Conference attendance and training through academic partnerships
    \item One-on-one mentoring for facility orientation
  \end{itemize}

  \fieldlabel{Prior experience:}
  \begin{itemize}
    \item Typically hired directly after PhD
    \item Internships or industry experience is helpful if individual does not have a PhD
  \end{itemize}
}
}

\clearpage
\newcommand{\ProfileHardwareTwenty}{%
\begin{tcolorbox}[
  enhanced, breakable, sharp corners,
  colback=hardwarelight,
  colframe=hardwaredark!60,
  boxrule=0.4pt,
  arc=2mm,
  borderline west={3pt}{0pt}{hardwaredark},
  left=10pt, right=10pt, top=8pt, bottom=8pt,
  title={\centering\sffamily\large\textcolor{white}{Profile H3.2: Quantum Hardware System Engineers}},
  fonttitle=\sffamily\large,
  coltitle=white,
]

\begin{center}
  {\scriptsize\textcolor{gray!}{\itshape Findings in this profile are not presented in order of importance nor does any individual position in the data include all listed elements.}}
\end{center}

\vspace{4pt}
{\small
\begin{itemize}[label={}, leftmargin=1.5em, itemsep=1pt, topsep=1pt]
  \item \textbf{Individual positions:} Test and Integration Engineer, Senior Research Scientist, Quantum Engineer
  \item \textbf{Company types:} Quantum computing hardware (platforms: trapped-ion, neutral-atom), Quantum algorithms and software, Quantum networking and communication
\end{itemize}
}

\ProfileSectionHeader[building]{hardwaredark}{Occupation-Specific Information}
\fieldlabel{Description:} Quantum hardware system engineers turn research into devices. They design hardware subsystems operating within practical constraints and integrate them into larger systems.

\vspace{0.5em}
\fieldlabel{Tasks:} On the job, they would…
\begin{itemize}
  \item Define hardware requirements for different systems
  \item Design hardware systems
  \item Generate bills of materials
  \item Integrate subsystems in a device
  \item Optimize hardware and software systems
  \item Test and calibrate hardware systems
  \item Work on productization of experimental results
\end{itemize}

\ProfileSectionHeader[brain]{hardwaredark}{Worker Requirements: Knowledge, Skills, Abilities (KSAs)}
\fieldlabel{Knowledge of:}
\begin{itemize}
  \item Control and RF (Radio Frequency) electronics
  \item Design software (e.g., SolidWorks, Onshape)
  \item Integrated systems that support trapped ions
  \item Optics
  \item Quantum mechanics 
  \item Software tools (e.g., Python, Conda, Jupyter)
  \item Systems engineering 
\end{itemize}

\fieldlabel{Occupation-specific skills and abilities:}
\begin{itemize}
  \item Able to conduct scientific research in an engineering focused environment
  \item Able to use programming skills for data collection and analysis
\end{itemize}

\fieldlabel{General skills and abilities:}
\begin{itemize}
  \item Communication, program management, and time management
\end{itemize}

\ProfileSectionHeader[graduation-cap]{hardwaredark}{Experience Requirements}
\fieldlabel{Education:} 
\begin{itemize}
\item Degrees: Bachelor, Master, or PhD
\item Disciplines: Electrical Engineering, Engineering, or Physics
\end{itemize}

\fieldlabel{On-the-job training:} 
\begin{itemize}
\item Training provided for job-specific tools and equipment
\end{itemize}

\fieldlabel{Prior experience:}
\begin{itemize}
  \item 1–2 years industry experience is typical
  \item PhD may suffice without prior industry experience
\end{itemize}

\end{tcolorbox}
}

\clearpage
\newcommand{\ProfileHardwareTwentyOne}{%
\begin{tcolorbox}[
  enhanced, breakable, sharp corners,
  colback=hardwarelight,
  colframe=hardwaredark!60,
  boxrule=0.4pt,
  arc=2mm,
  borderline west={3pt}{0pt}{hardwaredark},
  left=10pt, right=10pt, top=8pt, bottom=8pt,
  title={\centering\sffamily\large\textcolor{white}{Profile H3.3: Field Deployment Engineers}},
  fonttitle=\sffamily\large,
  coltitle=white,
]

\begin{center}
  {\scriptsize\textcolor{gray!}{\itshape Findings in this profile are not presented in order of importance and will be refined as additional interviews are conducted.}}
\end{center}

\vspace{4pt}
{\small
\begin{itemize}[label={}, leftmargin=1.5em, itemsep=1pt, topsep=1pt]
  \item \textbf{Individual positions:} Field Service Engineer 
  \item \textbf{Company types:} Quantum computing hardware, Enabling technologies
\end{itemize}
}

\ProfileSectionHeader[building]{hardwaredark}{Occupation-Specific Information}
\fieldlabel{Description:} Field deployment engineers deploy quantum technology products offsite and ensure the hardware operates as expected in the customer’s environment. Their main focus is to ensure the product is functioning correctly during offsite deployment.

\vspace{0.5em}
\fieldlabel{Tasks:} On the job, they would...
\begin{itemize}
  \item Deliver and install equipment at client sites
  \item Integrate equipment with client infrastructure
\end{itemize}

\ProfileSectionHeader[brain]{hardwaredark}{Worker Requirements: Knowledge, Skills, Abilities (KSAs)}
\fieldlabel{Occupation-specific skills and abilities:}
\begin{itemize}
    \item Able to adapt to changing client contexts
    \item Able to troubleshoot hardware
    \item Able to work on optical breadboards 
\end{itemize}

\fieldlabel{General skills and abilities:}
\begin{itemize}
  \item Communication, hands-on mechanical work, and problem solving
\end{itemize}

\ProfileSectionHeader[graduation-cap]{hardwaredark}{Experience Requirements}
\fieldlabel{Education:} 
\begin{itemize}
\item Degrees: Master or PhD
\item Disciplines: Chemistry, Electrical Engineering, or Physics
\end{itemize}

\fieldlabel{On-the-job training:} 
\begin{itemize}
\item Structured one-on-one training over 6–12 months
\end{itemize}

\fieldlabel{Prior experience:}
\begin{itemize}
  \item About half of candidates have similar prior roles
\end{itemize}

\end{tcolorbox}
}

\clearpage
\newcommand{\ProfileHardwareTwentyTwo}{%
\HardwareProfileBlock
{Profile H4.1: Superconducting Quantum Engineers}
{Quantum Hardware Engineer (x2)}
{Quantum computing hardware (platform: superconducting), Quantum algorithms and software, Enabling technologies, Consultants}
{%
  \fieldlabel{Description:} Superconducting quantum engineers model, design, fabricate, operate, and analyze superconducting qubits and circuits.

  \vspace{0.5em}
  \fieldlabel{Tasks: }On the job, they would...
\begin{itemize}
    \item Build and use quantum control systems
    \item Contribute to hardware development
    \item Design, develop, and fabricate qubits
    \item Design, operate, model, and analyze quantum circuits
    \item Develop low noise amplifiers
    \item Engage in microwave engineering
\end{itemize}

}
{%
  \fieldlabel{Knowledge of:}
\begin{itemize}
    \item Decoherence
    \item Electrical engineering 
    \item Josephson junctions
    \item Measurement processes
    \item Quantum design and control
    \item Statistics
    \item Superconducting qubits 
\end{itemize}

  \fieldlabel{Occupation-specific skills and abilities:}
  \begin{itemize}
    \item Able to do some programming in Python
  \end{itemize}

  \fieldlabel{General skills and abilities:}
  \begin{itemize}
    \item Data analysis, debugging, problem solving, project management, and troubleshooting
  \end{itemize}
}
{%
  \fieldlabel{Education:} 
  \begin{itemize}
  \item Degree: PhD 
  \item Disciplines: Engineering or Physics
  \end{itemize}

  \fieldlabel{On-the-job training:} 
  \begin{itemize}
  \item One-on-one mentoring 
  \item Workshops
  \end{itemize}

  \fieldlabel{Prior experience:} 
  \begin{itemize}
  \item Typically hired directly after PhD
  \end{itemize}
}
}

\clearpage
\newcommand{\ProfileHardwareTwentyThree}{%
\HardwareProfileBlock
{H4.2: Device Characterization \& Measurement Specialists}
{Quantum Metrology, Quantum Device Measurement Scientist, Lab Engineer}
{Quantum computing hardware (platform: superconducting), Quantum algorithms and software, Quantum sensing, Quantum networking and communication, Enabling technologies
}
{%
  \fieldlabel{Description:} Device characterization and measurement specialists perform device characterization by conducting specialized physical measurements (e.g., optical or  RF) of quantum systems and by performing analysis of the generated data. They use these to make determinations about device performance. 

  \vspace{0.5em}
  \fieldlabel{Tasks:} On the job, they would...
  \begin{itemize}
    \item Conduct physical measurements on quantum devices to understand their performance
      \item Coordinate multidisciplinary teams to ensure project specifications are met
    \item Write reports and publications
  \end{itemize}
}
{%
  \fieldlabel{Knowledge of:}
  \begin{itemize}
    \item Decoherence
    \item Quantum imaging techniques
    \item Quantum mechanics
  \end{itemize}

  \fieldlabel{Occupation-specific skills and abilities:}
\begin{itemize}
    \item Able to collaborate with fabrication and architecture teams
    \item Able to debug code
    \item Able to do some programming generally, without specifying whether it is for data processing, data representation, or other purposes
    \item Able to perform optical and RF (Radio Frequency) measurements
    \item Able to perform statistical analysis
    \item Able to troubleshoot hardware
\end{itemize}

}
{%
  \fieldlabel{Education:} 
  \begin{itemize}
  \item Degree: PhD
  \item Disciplines: Optics or Physics
  \end{itemize}

  \fieldlabel{On-the-job training:} 
  \begin{itemize}
  \item One-on-one mentoring 
  \item Workshops
  \end{itemize}

  \fieldlabel{Prior experience:}
  \begin{itemize}
     \item Hands-on experience with devices from PhD, internships, or industry is desirable
    \item Prior role at a metrology-focused company is preferred
    \item Typically hired directly after PhD
  \end{itemize}
}
}

\clearpage
\newcommand{\ProfileHardwareTwentyFour}{%
\begin{tcolorbox}[
  enhanced, breakable, sharp corners,
  colback=hardwarelight,
  colframe=hardwaredark!60,
  boxrule=0.4pt,
  arc=2mm,
  borderline west={3pt}{0pt}{hardwaredark},
  left=10pt, right=10pt, top=8pt, bottom=8pt,
  title={\centering\sffamily\large\textcolor{white}{Profile H4.3: Design Engineer, EE Circuits, RF Specialists}},
  fonttitle=\sffamily\large,
  coltitle=white,
]

\begin{center}
  {\scriptsize\textcolor{gray!}{\itshape Findings in this profile are not presented in order of importance and will be refined as additional interviews are conducted.}}
\end{center}

\vspace{4pt}
{\small
\begin{itemize}[label={}, leftmargin=1.5em, itemsep=1pt, topsep=1pt]
  \item \textbf{Individual positions:} Research Engineer
  \item \textbf{Company types:} Enabling technologies
\end{itemize}
}

\ProfileSectionHeader[building]{hardwaredark}{Occupation-Specific Information}
\fieldlabel{Description:}  Design engineer, EE (Electrical Engineering) circuits, RF (Radio Frequency) specialists develop and test classical electronic circuits for applications in quantum technologies. 

\vspace{0.5em}
\fieldlabel{Tasks:} On the job, they would…
\begin{itemize}
  \item Design and fabricate classical circuits 
  \item Design and fabricate jigs 
  \item Design and fabricate mechanical housings
\end{itemize}

\ProfileSectionHeader[brain]{hardwaredark}{Worker Requirements: Knowledge, Skills, Abilities (KSAs)}
\fieldlabel{Knowledge of:}
\begin{itemize}
    \item Diode lasers
    \item Quantum computing
    \item Quantum mechanics 
    \item Statistics
\end{itemize}

\fieldlabel{Occupation-specific skills and abilities:}
\begin{itemize}
  \item Able to communicate technical requirements for hardware to staff
  \item Able to do some programming generally, without specifying whether it is for data processing, data representation, or other purposes
  \item Able to make circuit design decisions based on other experimental factors (e.g., laser performance)
\end{itemize}

\fieldlabel{General skills and abilities:}
\begin{itemize}
  \item Communication, data analysis, modeling, and problem solving
\end{itemize}

\ProfileSectionHeader[graduation-cap]{hardwaredark}{Experience Requirements}
\fieldlabel{Education:} 
\begin{itemize}
\item Degrees: Bachelor or Master
\item Disciplines: Chemistry, Computer Science, Engineering, or Physics
\end{itemize}

\fieldlabel{On-the-job training:} 
\begin{itemize}
\item One-on-one mentoring 
\item External courses on specific optical techniques
\end{itemize}

\fieldlabel{Prior experience:} 
\begin{itemize}
\item Relevant work experience is helpful but not required
\end{itemize}

\end{tcolorbox}
}

\clearpage
\newcommand{\ProfileHardwareTwentyFive}{%
\begin{tcolorbox}[
  enhanced, breakable, sharp corners,
  colback=hardwarelight,
  colframe=hardwaredark!60,
  boxrule=0.4pt,
  arc=2mm,
  borderline west={3pt}{0pt}{hardwaredark},
  left=10pt, right=10pt, top=8pt, bottom=8pt,
  title={\centering\sffamily\large\textcolor{white}{Profile H4.5: Optics \& Photonics Assembly Specialists}},
  fonttitle=\sffamily\large,
  coltitle=white,
]

\begin{center}
  {\scriptsize\textcolor{gray!}{\itshape Findings in this profile are not presented in order of importance, nor does any individual position in the data include all listed elements.}}
\end{center}

\vspace{4pt}
{\small
\begin{itemize}[label={}, leftmargin=1.5em, itemsep=1pt, topsep=1pt]
  \item \textbf{Individual positions:} Fabrication Engineer, Assembly Technician, Optics Assembly Technician 
  \item \textbf{Company types:} Quantum computing hardware (platform: neutral-atom), Quantum algorithms and software, Quantum sensing, Quantum networking and communication, Enabling technologies
\end{itemize}
}

\ProfileSectionHeader[building]{hardwaredark}{Occupation-Specific Information}
\fieldlabel{Description:} Optics and photonics assembly specialists perform assembly, testing, and quality control for optical/photonic systems.

\vspace{0.5em}
\fieldlabel{Tasks:} On the job, they would...
\begin{itemize}
  \item Assemble and test optical systems 
  \item Clean and inspect optics for quality control
  \item Manage supply chain documentation (e.g.,  fill out certificates of compliance for products that leave the factory)
  \item Perform interferometry or microscopy measurements
  \item Splice fibers
\end{itemize}

\ProfileSectionHeader[brain]{hardwaredark}{Worker Requirements: Knowledge, Skills, Abilities (KSAs)}
\fieldlabel{Knowledge of:}
\begin{itemize}
    \item Electricity and magnetism 
    \item Non-quantum electronics
    \item Quantum information science
    \item Statistics
    \item Vacuum science
\end{itemize}

\fieldlabel{Occupation-specific skills and abilities:}
\begin{itemize}
  \item Able to debug code
  \item Able to interpret measurement results 
  \item Able to operate basic test and measurement devices (e.g., power supplies, multimeters)
  \item Able to use interferometers and microscopes
  \item Able to use lasers
  \item Able to work in a detail-oriented and cautious manner
\end{itemize}

\fieldlabel{General skills and abilities:}
\begin{itemize}
  \item Collaboration, communication, critical thinking, general experimental skills, problem solving, and project management
\end{itemize}

\ProfileSectionHeader[graduation-cap]{hardwaredark}{Experience Requirements}
\fieldlabel{Education:} 
\begin{itemize}
    \item Degrees: Associate or Bachelor
    \item Disciplines: Mathematics, Optics, Optical Engineering, or Physics
\end{itemize}

\fieldlabel{On-the-job training:} 
\begin{itemize}
\item Company talks
\item Hands-on training
\end{itemize}

\fieldlabel{Prior experience:} 
\begin{itemize}
\item 3-4 years of experience building, cleaning, handling, and inspecting optics and lasers is expected
\end{itemize}
\end{tcolorbox}
}

\clearpage
\newcommand{\ProfileHardwareTwentySix}{%
\HardwareProfileBlock
{Profile H4.4: Optics \& Photonics Experiment Specialists}
{Laser and Optics Engineer, Photonics Assembly Technician, Junior Photonics Experimenter}
{Quantum computing hardware (platform: trapped-ion), Quantum algorithms and software, Quantum sensing, Quantum networking and communication, Enabling technologies}
{%
  \fieldlabel{Description:} Optics and photonics experiment specialists support experiments, such as maintaining optical systems and performing data collection and analysis. 

  \vspace{0.5em}
  \fieldlabel{Tasks:} On the job, they would...
  \begin{itemize}
    \item Assist in running optics and photonics experiments
    \item Design and maintain optical systems
    \item Define hardware specifications and procure necessary equipment
    \item Set up lasers and beam lines
  \end{itemize}
}
{%
  \fieldlabel{Knowledge of:}
  \begin{itemize}
      \item Experimental optics
      \item Lab electronics 
      \item Statistics
  \end{itemize}

  \fieldlabel{Occupation-specific skills and abilities:}
  \begin{itemize}
    \item Able to troubleshoot setups
    \item Able to work with non-quantum electronics
    \item Able to work with optical experiments 
    \item Able to do some programming generally, without specifying whether it is for data processing, data representation, or other purposes
  \end{itemize}

  \fieldlabel{General skills and abilities:}
  \begin{itemize}
    \item Collaboration, communication, data analysis, debugging, and general experimental skills
  \end{itemize}
}
{%
  \fieldlabel{Education:}
  \begin{itemize}
  \item Degree: Bachelor
  \item Disciplines: Chemistry, Engineering, or Physics
  \end{itemize}

  \fieldlabel{On-the-job training:} 
  \begin{itemize}
  \item One-on-one mentoring 
  \item Workshops for specific hardware
  \end{itemize}

  \fieldlabel{Prior experience:} 
  \begin{itemize}
  \item Prior industry position or internships are helpful but not required
  \end{itemize} 
}
}

\clearpage
\newcommand{\ProfileHardwareTwentySeven}{%
\begin{tcolorbox}[
  enhanced, breakable, sharp corners,
  colback=hardwarelight,
  colframe=hardwaredark!60,
  boxrule=0.4pt,
  arc=2mm,
  borderline west={3pt}{0pt}{hardwaredark},
  left=10pt, right=10pt, top=8pt, bottom=8pt,
  title={\centering\sffamily\large\textcolor{white}{Profile H4.6: Cryogenics Specialists}},
  fonttitle=\sffamily\large,
  coltitle=white,
]

\begin{center}
  {\scriptsize\textcolor{gray!}{\itshape Findings in this profile are not presented in order of importance and will be refined as additional interviews are conducted.}}
\end{center}

\vspace{4pt}
{\small
\begin{itemize}[label={}, leftmargin=1.5em, itemsep=1pt, topsep=1pt]
  \item \textbf{Individual positions:} Cryogenic Physicist 
  \item \textbf{Company types:} Quantum computing hardware, Quantum algorithms and software, Quantum sensing, Quantum networking and communication, Enabling technologies 
\end{itemize}
}

\ProfileSectionHeader[building]{hardwaredark}{Occupation-Specific Information}
\fieldlabel{Description:} Cryogenics specialists develop systems to test quantum devices at cryogenic temperatures.

\vspace{0.5em}
\fieldlabel{Tasks:} On the job, they would...
\begin{itemize}
  \item Develop cryogenic test systems
\end{itemize}

\ProfileSectionHeader[brain]{hardwaredark}{Worker Requirements: Knowledge, Skills, Abilities (KSAs)}
\fieldlabel{Knowledge of:}
\begin{itemize}
  \item Cryogenics 
  \item Detector and sensing platforms
  \item Statistics
\end{itemize}

\fieldlabel{Occupation-specific skills and abilities:}
\begin{itemize}
    \item Able to program and debug code, and troubleshoot experiments
\end{itemize}

\fieldlabel{General skills and abilities:}
\begin{itemize}
  \item Adaptability, collaboration, communication, critical thinking, data analysis, modeling, problem solving, and project management
\end{itemize}

\ProfileSectionHeader[graduation-cap]{hardwaredark}{Experience Requirements}
\fieldlabel{Education:} 
\begin{itemize}
\item Degree: PhD 
\item Disciplines: Engineering or Physics
\end{itemize}

\fieldlabel{On-the-job training:} 
\begin{itemize}
\item One-on-one mentoring
\end{itemize}

\fieldlabel{Prior experience:} 
\begin{itemize}
\item None required
\end{itemize}

\end{tcolorbox}
}

\clearpage
\newcommand{\ProfileHardwareTwentyEight}{%
\HardwareProfileBlock
{Profile H4.7: Nano/Microscale Specialists}
{Patterning Integration Engineer, Nanofabrication Engineer}
{Quantum computing hardware (platform: superconducting), Quantum networking and communication, Enabling technologies}
{%
  \fieldlabel{Description:} Nano/microscale specialists fabricate nano/microscale devices, typically in a cleanroom setting. They conduct nano/microscale patterning, utilizing various techniques dependent on the architecture of the devices. They perform characterization of fabricated nano/microscale devices.

  \vspace{0.5em}
  \fieldlabel{Tasks:} On the job, they would...
\begin{itemize}
    \item Characterize and package integrated photonic chips
    \item Combine traditional manufacturing techniques with quantum devices
    \item Fabricate integrated photonic chips
    \item Refine fabrication techniques based on feedback
    \item Work in a cleanroom setting
\end{itemize}

}
{%
  \fieldlabel{Knowledge of:}
\begin{itemize}
    \item Cleanroom optical fabrication processes
    \item Function of quantum emitters
    \item Quantum device design
    \item Quantum hardware
    \item Quantum technology implementation
\end{itemize}

  \fieldlabel{Occupation-specific skills and abilities:}
\begin{itemize}
    \item Able to adhere to cleanroom protocols
    \item Able to do some programming generally, without specifying whether it is for data processing, data representation, or other purposes
\end{itemize}
  \fieldlabel{General skills and abilities:}
  \begin{itemize}
    \item Adaptability, collaboration, communication, critical thinking, problem solving, and project management
  \end{itemize}
}
{%
  \fieldlabel{Education:} 
  \begin{itemize}
  \item Degrees: Master or PhD 
  \item Disciplines: Engineering, Material Science, or Physics
  \end{itemize}

  \fieldlabel{On-the-job training:} 
  \begin{itemize}
  \item Conferences
  \item External courses
  \item One-on-one mentoring
  \item Workshops
  \end{itemize}

  \fieldlabel{Prior experience:} 
  \begin{itemize}
  \item PhD is usually sufficient
  \item Some industry experience in fabrication is helpful
  \end{itemize} 
}
}

\clearpage
\newcommand{\ProfileHardwareTwentyNine}{%
\HardwareProfileBlock
{Profile H4.8: Lab \& Construction Technicians}
{Lab Technician, Construction Specialist}
{Quantum computing hardware (platform: trapped-ion), Quantum algorithms and software, Enabling technologies 
}
{%
  \fieldlabel{Description:} Lab \& construction technicians build and maintain facilities for the fabrication and testing of quantum devices. This includes planning and overseeing the construction of cleanroom facilities, maintaining mechanical connections, and general electrical or plumbing work.

  \vspace{0.5em}
  \fieldlabel{Tasks:} On the job, the would...
\begin{itemize}
    \item Check hardware placement and installation
    \item Coordinate cleanroom construction projects for quantum facility 
    \item Maintain mechanical connections
    \item Perform electrical and plumbing work
    \item Perform general lab maintenance 
\end{itemize}

}
{%
  \fieldlabel{Occupation-specific skills and abilities:}
\begin{itemize}
    \item Able to do some programming generally, without specifying whether it is for data processing, data representation, or other purposes
    \item Able to read and draw architectural plans
    \item Able to work in chip fabrication facilities
\end{itemize}

  \fieldlabel{General skills and abilities:}
  \begin{itemize}
    \item Communication, critical thinking, leadership, problem solving, and project management
  \end{itemize}
}
{%
  \fieldlabel{Education:} 
  \begin{itemize}
  \item Degrees: Associate or higher
  \item Disciplines: Engineering or Physics
  \end{itemize}

  \fieldlabel{On-the-job training:} 
\begin{itemize}
  \item External courses
  \item One-on-one mentoring
\end{itemize}

  \fieldlabel{Prior experience:} 
  \begin{itemize}
  \item Typically some prior lab construction experience is expected
\end{itemize}
}
}

\section*{Hardware Roles Profiles}
\addcontentsline{toc}{section}{ Hardware Roles Profiles}
The hardware category consists of roles focused on designing, building, maintaining, and scaling quantum hardware systems. Within the hardware category, roles were grouped by leadership responsibilities and the breadth of technical expertise (e.g., does the role require specialization for a specific subsystem or expertise spanning several subsystems and their integration).
\subsection*{H1: Technical Managers}
\addcontentsline{toc}{subsection}{H1: Technical Managers}
Technical managers fill a dual role, bringing significant technical expertise and the ability to manage and coordinate teams for a concentrated hardware development effort. The two roles focused on managing people and coordinating hardware development efforts. The breath of responsibilities requires broad technical knowledge.

\subsubsection*{H1.1 Senior Scientists}
\addcontentsline{toc}{subsubsection}{H1.1 Senior Scientists}
\ProfileHardwareFifteen
\clearpage

\subsubsection*{H1.2 Engineering Managers}
\addcontentsline{toc}{subsubsection}{H1.2 Engineering Managers}
\ProfileHardwareSixteen
\clearpage

\subsection*{H2: Technical Leads}
\addcontentsline{toc}{subsection}{H2: Technical Leads}

Similar to the technical managers, the technical leads bring significant technical expertise and the ability to manage and lead different hardware development efforts, but at a smaller scale than the technical manager. While  technical managers often coordinate multiple teams, technical leads typically focus on driving specific experiments or product development efforts for a single team. 

\subsubsection*{H2.1 Commercialization Leads}
\addcontentsline{toc}{subsubsection}{H2.1 Commercialization Leads}
\ProfileHardwareSeventeen
\clearpage
\subsubsection*{H2.2 Systems Engineers}
\addcontentsline{toc}{subsubsection}{H2.2 Systems Engineers}
\ProfileHardwareEighteen
\clearpage

\subsection*{H3: Technical System Specialists}
\addcontentsline{toc}{subsection}{H3: Technical System Specialists}

Technical system specialists hold domain expertise of hardware systems and are able to work with multiple subsystems necessary to develop different quantum technologies. 

\subsubsection*{H3.1 Experimental Scientists}
\addcontentsline{toc}{subsubsection}{H3.1 Experimental Scientists}
\ProfileHardwareNineteen
\clearpage
\subsubsection*{H3.2 Quantum Hardware System Engineers}
\addcontentsline{toc}{subsubsection}{H3.2 Quantum Hardware System Engineers}
\ProfileHardwareTwenty
\clearpage

\subsubsection*{H3.3 Field Deployment Engineers}
\addcontentsline{toc}{subsubsection}{H3.3 Field Deployment Engineers}
\ProfileHardwareTwentyOne
\clearpage

\subsection*{H4: Technical Subsystem Specialists}
\addcontentsline{toc}{subsection}{H4: Technical  Subsystem Specialists}

Technical subsystem specialists are roles that have a specific expertise or skill-set related to a subsystem needed for quantum technology development or manufacturing.
\subsubsection*{H4.1 Superconducting Quantum Engineers}
\addcontentsline{toc}{subsubsection}{H4.1 Superconducting Quantum Engineers}
\ProfileHardwareTwentyTwo
\clearpage

\subsubsection*{H4.2 Device Characterization \& Measurement Specialists}
\addcontentsline{toc}{subsubsection}{H4.2 Device Characterization \& Measurement Specialists}
\ProfileHardwareTwentyThree
\clearpage

\subsubsection*{H4.3 Design Engineer, EE Circuits, RF
Specialists}
\addcontentsline{toc}{subsubsection}{H4.3 Design Engineer, EE Circuits, RF
Specialists}
\ProfileHardwareTwentyFour
\clearpage

\subsubsection*{H4.4 Optics \& Photonics Experiment Specialists}
\addcontentsline{toc}{subsubsection}{H4.4 Optics \& Photonics Experiment Specialists}
\ProfileHardwareTwentySix
\clearpage

\subsubsection*{H4.5 Optics \& Photonics Assembly Specialists}
\addcontentsline{toc}{subsubsection}{H4.5 Optics \& Photonics Assembly Specialists}
\ProfileHardwareTwentyFive
\clearpage

\subsubsection*{H4.6 Cryogenics Specialists}
\addcontentsline{toc}{subsubsection}{H4.6 Cryogenics Specialists}
\ProfileHardwareTwentySeven
\clearpage

\subsubsection*{H4.7 Nano/Microscale Specialists}
\addcontentsline{toc}{subsubsection}{H4.7 Nano/Microscale Specialists}
\ProfileHardwareTwentyEight
\clearpage

\subsubsection*{H4.8 Lab \& Construction Technicians}
\addcontentsline{toc}{subsubsection}{H4.8 Lab \& Construction Technicians}
\ProfileHardwareTwentyNine

\clearpage

\clearpage
\section*{Software Roles Profiles}
\addcontentsline{toc}{section}{Software Roles Profiles}

The software category consists of roles focused on designing, developing, and optimizing software for quantum systems and applications.  Within the software category, there are roles more focused on developing new software utilizing primarily classical methods and roles focused on developing algorithms for, or running algorithms on, quantum hardware. 

\newcommand{\ProfileSoftwareEight}{%
\begin{tcolorbox}[
  enhanced, breakable, sharp corners,
  colback=softwaredark!5, colframe=softwaredark!70,
  boxrule=0.4pt, arc=2mm,
  borderline west={3pt}{0pt}{softwaredark},
  left=10pt, right=10pt, top=8pt, bottom=8pt,
  title={\centering\sffamily\large\textcolor{white}{Profile S1.1: Traditional Software Engineers}},
  fonttitle=\sffamily\large,
  coltitle=white,
]
\begin{center}
{\scriptsize\textcolor{gray!}{\itshape Findings in this profile are not presented in order of importance, nor does any individual position in the data include all listed elements.}}
\end{center}
\vspace{4pt}
{\small
\begin{itemize}[label={}, leftmargin=1.5em, itemsep=1pt, topsep=1pt]
  \item \textbf{Individual positions:} Software Developer, Software Engineer, Front End Developer, Junior Scientific Software Engineer, Senior Scientific Software Engineer 
  \item \textbf{Company types:} Quantum computing hardware (platform: neutral-atom), Quantum algorithms and software, Enabling technologies, Consultants
\end{itemize}
}

\ProfileSectionHeader[building]{softwaredark}{Occupation-Specific Information}
\fieldlabel{Description:} Traditional software engineers build classical software systems that support or integrate with quantum workflows.

\fieldlabel{Tasks:}
\begin{itemize}
    \item Create chemistry modeling interfaces and biomedical dashboards
    \item Design and implement Application Programming Interfaces (APIs), web interfaces, and other software stack components for delivering quantum systems
    \item Develop cloud delivery systems that allow users to submit jobs to be routed into quantum computers, and return results
    \item Develop robust and reliable cloud applications
    \item Innovate on top of existing software stacks to enhance platform capabilities
    \item Integrate components of the quantum cloud service platform 
    \item Maintain production and test systems for software
\end{itemize}

\ProfileSectionHeader[brain]{softwaredark}{Worker Requirements: Knowledge, Skills, Abilities (KSAs)}
\fieldlabel{Knowledge of:}
\begin{itemize}
    \item Programming 
    \item Quantum chemistry representations
    \item Quantum circuits including their properties are and how they are executed 
     \item Software development
\end{itemize}

\fieldlabel{Occupation-specific skills:}
\begin{itemize}
  \item Able to do some programming generally, without specifying whether it is for data processing, data representation, or other purposes
\end{itemize}
\fieldlabel{General skills:}
\begin{itemize}
  \item Adaptability, collaboration, communication, critical thinking, data analysis, problem solving, and troubleshooting
\end{itemize}

\ProfileSectionHeader[graduation-cap]{softwaredark}{Experience Requirements}
\fieldlabel{Education:} 
\begin{itemize}
  \item Degrees: Bachelor, Master, or PhD
  \item Disciplines: Chemistry, Computer Science, Engineering, or Mathematics
\end{itemize} 

\fieldlabel{On-the-job training:}
\begin{itemize}
  \item External courses
  \item Internal courses for specific company knowledge
  \item One-on-one mentoring
\end{itemize}
\fieldlabel{Prior experience:} 
\begin{itemize}
  \item Internship experience can be helpful but not required
  \item Strong coding and software development skills are what is needed for consideration
\end{itemize}
\end{tcolorbox}
}

\newcommand{\ProfileSoftwareNine}{%
\begin{tcolorbox}[
  enhanced, breakable, sharp corners,
  colback=softwaredark!5, colframe=softwaredark!70,
  boxrule=0.4pt, arc=2mm,
  borderline west={3pt}{0pt}{softwaredark},
  left=10pt, right=10pt, top=8pt, bottom=8pt,
  title={\centering\sffamily\large\textcolor{white}{Profile S1.2: Quantum Software Engineers}},
  fonttitle=\sffamily\large,
  coltitle=white,
]
\begin{center}
  {\scriptsize\textcolor{gray!}{\itshape Findings in this profile are not presented in order of importance, nor does any individual position in the data include all listed elements.}}
\end{center}
\vspace{4pt}
{\small
\begin{itemize}[label={}, leftmargin=1.5em, itemsep=1pt, topsep=1pt]
  \item \textbf{Individual positions:} Senior Quantum Application Engineer, Quantum Computational Scientist, Software Developer, Quantum Software Developer, Quantum Software Engineer
  \item \textbf{Company types:} Quantum computing hardware (platform: neutral-atom, superconducting), Quantum algorithms and software, Quantum sensing, Quantum networking and communication, Enabling technologies, Consultants
\end{itemize}
}
\ProfileSectionHeader[building]{softwaredark}{Occupation-Specific Information}
\fieldlabel{Description:} Quantum software engineers design and develop software directly related to quantum computing.

\fieldlabel{Tasks:} On the job, they would...
\begin{itemize}
    \item Develop and integrate classical and quantum software
    \item Develop research code into shareable packages 
    \item Ensure compatibility of code across different computing environments where the software will be used
    \item Plan and organize quantum software development projects
    \item Read, write, and present research
    \item Translate domain knowledge into quantum algorithms to bridge technical and application areas
    \item Write code for quantum applications, which includes qubit control,  web app development, and language design
\end{itemize}

\ProfileSectionHeader[brain]{softwaredark}{Worker Requirements: Knowledge, Skills, Abilities (KSAs)}
\fieldlabel{Knowledge of:}
\begin{itemize}
    \item Classical computation and high performance computing (HPC) centric languages
    \item Cloud platforms used in quantum computing environments 
    \item Data science
    \item Foundational knowledge of AMO physics 
    \item General programming tools (e.g., Python, Rust, and Git)
    \item Quantum algorithms and their applications
    \item Quantum circuits, including mathematical operations, the measurement process, normal operators in quantum, and expectation values
    \item Quantum computing theory and its practical challenges
    \item Software package design (SPACK)
\end{itemize}

\fieldlabel{Occupation-specific skills and abilities:}
\begin{itemize}
  \item Able to read and write scientific papers 
  \item Able to work with ambiguity 
\end{itemize}
\fieldlabel{General skills:}
\begin{itemize}
  \item Adaptability, critical thinking, documentation, problem solving, and project management
\end{itemize}

\ProfileSectionHeader[graduation-cap]{softwaredark}{Experience Requirements}
\fieldlabel{Education:} 
\begin{itemize}
  \item Degrees: Bachelor, Master or PhD
  \item Disciplines: Chemistry, Computer Science, Engineering, or Mathematics
\end{itemize}

\fieldlabel{On-the-job training:} 
\begin{itemize}
  \item One-on-one mentoring
  \item Participation in external courses
\end{itemize}

\fieldlabel{Prior experience:} 
\begin{itemize}
\item Experience at another company or an industry internship is preferred
\end{itemize}
\end{tcolorbox}
}

\newcommand{\ProfileSoftwareTen}{%
\begin{tcolorbox}[
  enhanced, breakable, sharp corners,
  colback=softwaredark!5, colframe=softwaredark!70,
  boxrule=0.4pt, arc=2mm,
  borderline west={3pt}{0pt}{softwaredark},
  left=10pt, right=10pt, top=8pt, bottom=8pt,
  title={\centering\sffamily\large\textcolor{white}{Profile S2.1: Quantum Information Science Algorithms Theorists}},
  fonttitle=\sffamily\large,
  coltitle=white,
]
\begin{center}
  {\scriptsize\textcolor{gray!}{\itshape Findings in this profile are not presented in order of importance and will be refined as additional interviews are conducted.}}
\end{center}
\vspace{4pt}
{\small
\begin{itemize}[label={}, leftmargin=1.5em, itemsep=1pt, topsep=1pt]
  \item \textbf{Individual positions:} Quantum Theory Algorithms Researcher
  \item \textbf{Company types:} Quantum computing hardware (platform: superconducting), Quantum algorithms and software
\end{itemize}
}

\ProfileSectionHeader[building]{softwaredark}{Occupation-Specific Information}
\fieldlabel{Description:} Quantum information science algorithms theorists utilize theory for conceptual
design, mathematical formulation, and optimization of quantum algorithms.

\fieldlabel{Tasks:}
\begin{itemize}
  \item Apply advanced mathematical methods to analyze, optimize, and validate quantum algorithms' performance 
  \item Engage in white-boarding sessions to explore, derive, and visualize new algorithmic ideas and theoretical approaches 
\end{itemize}

\ProfileSectionHeader[brain]{softwaredark}{Worker Requirements: Knowledge, Skills, Abilities (KSAs)}
\fieldlabel{Knowledge of:}
\begin{itemize}
    \item Classical algorithms and computational techniques
    \item Theoretical foundations and frameworks of quantum computation 
\end{itemize}
\fieldlabel{Skills:}
\begin{itemize}
  \item Able to compare classical computational techniques with quantum approach
  \item Able to do some programming generally, without specifying whether it is for data processing, data representation, or other purposes
\end{itemize}
\fieldlabel{General skills:}
\begin{itemize}
  \item Collaboration and communication
\end{itemize}

\ProfileSectionHeader[graduation-cap]{softwaredark}{Experience Requirements}
\fieldlabel{Education:}
\begin{itemize}
  \item Degree: PhD
  \item Disciplines: Computer Science or Physics
\end{itemize}

\fieldlabel{Training:} 
\begin{itemize}
  \item None
\end{itemize}

\fieldlabel{Prior experience:} 
\begin{itemize}
\item Industry exposure is an advantage 
\item  Internship experience can be beneficial but not required
\item Relevant academic or research experience is expected
\end{itemize}
\end{tcolorbox}
}

\newcommand{\ProfileSoftwareTwelve}{%
\begin{tcolorbox}[
  enhanced, breakable, sharp corners,
  colback=softwaredark!5, colframe=softwaredark!70,
  boxrule=0.4pt, arc=2mm,
  borderline west={3pt}{0pt}{softwaredark},
  left=10pt, right=10pt, top=8pt, bottom=8pt,
  title={\centering\sffamily\large\textcolor{white}{Profile S2.2: Quantum Algorithms Programmers}},
  fonttitle=\sffamily\large,
  coltitle=white,
]
\begin{center}
  {\scriptsize\textcolor{gray!}{\itshape Findings in this profile are not presented in order of importance, nor does any individual position in the data include all listed elements.}}
\end{center}
\vspace{4pt}
{\small
\begin{itemize}[label={}, leftmargin=1.5em, itemsep=1pt, topsep=1pt]
  \item \textbf{Individual positions:} Algorithm Developer, Research Scientist/Algorithm Developer
  \item \textbf{Company types:}  Quantum computing hardware, Quantum algorithms and software, Consultants, Quantum sensing, Quantum networking and communication, Enabling Technologies
\end{itemize}
}

\ProfileSectionHeader[building]{softwaredark}{Occupation-Specific Information}
\fieldlabel{Description:} Quantum algorithms programmers are focused on implementing, testing, and optimizing quantum algorithms on quantum computing platforms.

\fieldlabel{Tasks:}
\begin{itemize}
  \item Apply theoretical principles to develop and test quantum algorithms 
  \item Design and implement algorithms for quantum computing platforms
  \item Engage in research activities, which include reading and analyzing research papers 
  \item Write, test, and maintain code that supports experiment implementation 
\end{itemize}

\ProfileSectionHeader[brain]{softwaredark}{Worker Requirements: Knowledge, Skills, Abilities (KSAs)}
\fieldlabel{Knowledge of:}
\begin{itemize}
  \item Decoherence
  \item Error correction
  \item Linear algebra and statistics
  \item Open systems dynamics
  \item Quantum cryptography
  \item Quantum design and control 
  \item Quantum gates 
  \item Quantum imaging
  \item Quantum software
  \item Quantum technology implementation
\end{itemize}

\fieldlabel{Occupation-specific skills and abilities:}
\begin{itemize}
  \item Able to do some programming generally, without specifying whether it is for data processing, data representation, or other purposes

\end{itemize}
\fieldlabel{General skills and abilities:}
\begin{itemize}
  \item Adaptability, collaboration, communication, critical thinking, debugging, problem solving, and troubleshooting
\end{itemize}

\ProfileSectionHeader[graduation-cap]{softwaredark}{Experience Requirements}
\fieldlabel{Education:} 
\begin{itemize}
  \item Degrees: Bachelor or Master 
  \item Disciplines: Computer Science, Mathematics, or Physics
\end{itemize} 

\fieldlabel{On-the-job training:}
\begin{itemize}
  \item One-on-one mentoring
  \item Participation in workshops 
  \item Participation in online learning modules developed internally within a company 
  \item Participation in external courses 
\end{itemize}
\fieldlabel{Prior experience:} 
\begin{itemize}
\item Internship or prior company position is preferred
\end{itemize}
\end{tcolorbox}
}

\subsection*{S1: Software Engineering}
\addcontentsline{toc}{subsection}{S1: Software Engineering}
Software engineering roles research, design, and develop computer and cloud software or specialized programs. The dividing line between the following two roles are the focus and
application of those programs.
\subsubsection*{S1.1 Traditional Software Engineers}
\addcontentsline{toc}{subsubsection}{S1.1 Traditional Software Engineers}
\ProfileSoftwareEight
\clearpage
\subsubsection*{S1.2 Quantum Software Engineers}
\addcontentsline{toc}{subsubsection}{S1.2 Quantum Software Engineers}
\ProfileSoftwareNine
\clearpage

\subsection*{S2: Applications \& Algorithms}
\addcontentsline{toc}{subsection}{S2: Applications \& Algorithms}

Applications and algorithms specialists develop, implement, and optimize algorithms for quantum applications that run on quantum computers in order to solve relevant problems.

\subsubsection*{S2.1 Quantum Information Science Algorithms Theorists}
\addcontentsline{toc}{subsubsection}{S2.1 Quantum Information Science Algorithms Theorists}
\ProfileSoftwareTen
\clearpage
\subsubsection*{S2.2 Quantum Algorithms Programmers
}
\addcontentsline{toc}{subsubsection}{S2.2 Quantum Algorithms Programmers
}
\ProfileSoftwareTwelve


\clearpage
\section*{Bridging Roles Profiles}
\addcontentsline{toc}{section}{Bridging Roles Profiles}

The bridging category consists of roles that facilitate communication and collaboration between different roles within a company. Bridging roles require expertise in more than one domain for the purpose of ``bridging the gap'' between those domains to achieve a goal.

\newcommand{\ProfileBridgingSeven}{%
\begin{tcolorbox}[
  enhanced, breakable, sharp corners,
  colback=bridgingdark!5, colframe=bridgingdark!70,
  boxrule=0.4pt, arc=2mm,
  borderline west={3pt}{0pt}{bridgingdark},
  left=10pt, right=10pt, top=8pt, bottom=8pt,
  title={\centering\sffamily\large\textcolor{white}{Profile B1.2: Quantum Technology End Users}},
  fonttitle=\sffamily\large,
  coltitle=white,
]
\begin{center}
  {\scriptsize\textcolor{gray!}{\itshape Findings in this profile are not presented in order of importance and will be refined as additional interviews are conducted.}}
\end{center}
\vspace{4pt}
{\small
\begin{itemize}[label={}, leftmargin=1.5em, itemsep=1pt, topsep=1pt]
  \item \textbf{Individual positions:} Quantum Computational Scientist
  \item \textbf{Company types:} End users of quantum technology
\end{itemize}
}

\ProfileSectionHeader[building]{bridgingdark}{Occupation-Specific Information}
\fieldlabel{Description:} Quantum technology end users explore ways to leverage quantum technologies to solve domain-specific problems for their company, whose primary business is outside of QISE.

\fieldlabel{Tasks:} On the job, they would...
\begin{itemize}
  \item Communicate complex technical findings in non-technical language to stakeholders and decision makers
  \item Conduct broad research to explore the current state of the art in quantum computing,  which includes regularly scanning paper headlines, reading abstracts, and selecting paper for deeper review 
  \item Evaluate feasibility and relevance of potential quantum applications
  \item Identify application areas within the organization that could benefit from quantum technologies
\end{itemize}

\ProfileSectionHeader[brain]{bridgingdark}{Worker Requirements: Knowledge, Skills, Abilities (KSAs)}
\fieldlabel{Knowledge of:}
\begin{itemize}
  \item Big-picture understanding of present and future quantum computing and applications
  \item General quantum mechanics
  \item Linear algebra
  \item Quantum information theory
\end{itemize}

\fieldlabel{Occupation-specific skills and abilities:}
\begin{itemize}
  \item Able to do some programming generally, without specifying whether it is for data processing, data representation, or other purposes
  \item Able to explore and review current research
  \item Able write research papers
\end{itemize}

\fieldlabel{General skills and abilities:}
\begin{itemize}
  \item Collaboration, communication, and storytelling
\end{itemize}

\ProfileSectionHeader[graduation-cap]{bridgingdark}{Experience Requirements}
\fieldlabel{Education:}
\begin{itemize}
  \item Degree: PhD
  \item Discipline: Mathematics 
\end{itemize}

\fieldlabel{On-the-job training:}
\begin{itemize}
  \item Attending conferences and workshops
\end{itemize}

\fieldlabel{Prior experience:}
\begin{itemize}
  \item Not specified
\end{itemize}
\end{tcolorbox}
}

\newcommand{\ProfileBridgingEleven}{%
\begin{tcolorbox}[
  enhanced, breakable, sharp corners,
  colback=bridgingdark!5, colframe=bridgingdark!70,
  boxrule=0.4pt, arc=2mm,
  borderline west={3pt}{0pt}{bridgingdark},
  left=10pt, right=10pt, top=8pt, bottom=8pt,
  title={\centering\sffamily\large\textcolor{white}{Profile B1.1: Quantum Software Application Developer \& Trainers}},
  fonttitle=\sffamily\large,
  coltitle=white,
]
\begin{center}
  {\scriptsize\textcolor{gray!}{\itshape Findings in this profile are not presented in order of importance, nor does any individual position in the data include all listed elements.}}
\end{center}
\vspace{4pt}
{\small
\begin{itemize}[label={}, leftmargin=1.5em, itemsep=1pt, topsep=1pt]
  \item \textbf{Individual positions:} Quantum Software Engineer, Quantum Application Scientist/Applied Scientist, Back-End Developer
  \item \textbf{Company types:}  Quantum computing hardware (platform: superconducting), Quantum algorithms and software, Enabling technologies, Consultants
\end{itemize}
}

\ProfileSectionHeader[building]{bridgingdark}{Occupation-Specific Information}
\fieldlabel{Description:} Quantum software application developers \& trainers develop real-world applications that make use of quantum software and apply quantum algorithms. They also play a key role in training customers to effectively use quantum software tools.

\fieldlabel{Tasks:} On the job, they would...
\begin{itemize}
  \item Access and manage multiple cloud platforms
  \item Communicate technical quantum concepts clearly to non-technical team members
  \item Conduct research
  \item Contribute to open-source software projects
  \item Create and write educational materials
  \item Deliver lectures and host workshops
  \item Design and develop quantum software in specific application areas (e.g., optimization, topology)
  \item Provide scientific guidance on quantum applications
  \item Work with QPU (Quantum Processing Unit) cloud platforms (e.g., AWS Braket, Azure Quantum, IBM Quantum)
  \item Work with quantum sensor platforms and integrate them with other systems
  \item Write and maintain code to support quantum software development
\end{itemize}

\ProfileSectionHeader[brain]{bridgingdark}{Worker Requirements: Knowledge, Skills, Abilities (KSAs)}
\fieldlabel{Knowledge of:}
\begin{itemize}
  \item Capabilities and limitations of current quantum computers
  \item Code editors and common development environments
  \item Data structures, classical algorithms, and algorithm optimization
  \item Decoherence
  \item Domain application areas (e.g., chemistry simulations) and project-relevant knowledge
  \item Error correction
  \item Linear algebra and statistics
  \item Open systems dynamics
  \item Open-source conventions and workflows
  \item Programming languages, and software development practices
  \item Qubit hardware
  \item Quantum algorithms
  \item Quantum computing libraries for software development and hardware integration
  \item Quantum cryptography
  \item Quantum design and control
  \item Quantum imaging
  \item Quantum information science
  \item Quantum systems
  \item Quantum technology implementation
  \item System noise, error mitigation techniques and cost estimation for running quantum computing jobs
\end{itemize}

\fieldlabel{Occupation-specific skills and abilities:}
\begin{itemize}
  \item Able to do some programming generally, including working with Python, Git, and Github
  \item Able to implement abstract quantum algorithms efficiently on real hardware
\end{itemize}

\fieldlabel{General skills and abilities:}
\begin{itemize}
  \item Adaptability, collaboration, communication, critical thinking, data science, problem solving, project management, public speaking, troubleshooting, and writing
\end{itemize}

\ProfileSectionHeader[graduation-cap]{bridgingdark}{Experience Requirements}
\fieldlabel{Education:} 
\begin{itemize}
  \item Degrees: Bachelor, Master, or PhD
  \item Disciplines: Chemistry, Computer Science, or Physics
\end{itemize} 

\fieldlabel{On-the-job training:}
\begin{itemize}
  \item External courses and online learning  primarily focused on filling knowledge and skills gaps
  \item One-on-one mentorship
\end{itemize}

\fieldlabel{Prior experience:}
\begin{itemize}
  \item Candidates may be hired directly from PhD programs
  \item Internships at other companies may be helpful but are not always required
  \item Relevant academic or research experience is expected
\end{itemize}
\end{tcolorbox}
}

\newcommand{\ProfileBridgingThirteen}{%
\begin{tcolorbox}[
  enhanced, breakable, sharp corners,
  colback=bridgingdark!5, colframe=bridgingdark!70,
  boxrule=0.4pt, arc=2mm,
  borderline west={3pt}{0pt}{bridgingdark},
  left=10pt, right=10pt, top=8pt, bottom=8pt,
  title={\centering\sffamily\large\textcolor{white}{Profile B2.1: Device \& System Hardware Computational Scientists}},
  fonttitle=\sffamily\large,
  coltitle=white,
]
\begin{center}
  {\scriptsize\textcolor{gray!}{\itshape Findings in this profile are not presented in order of importance, nor does any individual position in the data include all listed elements.}}
\end{center}
\vspace{4pt}
{\small
\begin{itemize}[label={}, leftmargin=1.5em, itemsep=1pt, topsep=1pt]
  \item \textbf{Individual positions:} AMO Theorist, Quantum Theoretical Scientist, Photonics Engineer, Computational RF Scientist, Quantum Research Scientist, Quantum Error Correction Scientist
  \item \textbf{Company types:} Quantum computing hardware (platform: trapped-ion, neutral-atom), Quantum algorithms and software, Quantum sensing, Quantum networking and communication, Enabling technologies, Consultants
\end{itemize}
}

\ProfileSectionHeader[building]{bridgingdark}{Occupation-Specific Information}
\fieldlabel{Description:} Device \& system hardware computational scientists computationally model quantum devices and systems to support design, validation, and optimization.

\fieldlabel{Tasks:} On the job, they would...
\begin{itemize}
    \item Model and analyze real-world systems to identify performance issues and determine theoretical adjustments or control parameters that lead to actionable results
    \item Model and simulate dynamics in qubit devices
    \item Perform benchmarking studies primarily through numerical simulations
    \item Perform literature reviews to identify methods, theories, or results to improve ongoing experiments 
    \item Perform quantum error correction simulations
    \item Simulate light behavior in optical cavities
\end{itemize}

\ProfileSectionHeader[brain]{bridgingdark}{Worker Requirements: Knowledge, Skills, Abilities (KSAs)}
\fieldlabel{Knowledge of:}
\begin{itemize}
  \item Device physics
  \item Large scale and efficient simulation, programming, and debugging of physical systems 
  \item Noise models 
  \item Open systems theory
  \item Process matrices
  \item Quantum error correction theory
  \item Quantum software
  \item Randomized benchmarking theory
  \item Stabilizer codes
  \item Statistics
  \item System integration principles and hardware operation
  \item Theoretical quantum information science grounded in physical implementations
\end{itemize}

\fieldlabel{Occupation-specific skills and abilities:}
\begin{itemize}
  \item Able to do some programming generally, without specifying whether it is for data processing, data representation, or other purposes
  \item Able to effectively collaborate with system integration engineers 
  \item Able to perform numerical and physical simulations of quantum and optical systems 
  \item Able to use AI (Artificial Intelligence) tools to assist in literature review
\end{itemize}

\fieldlabel{General skills and abilities:}
\begin{itemize}
  \item Adaptability, collaboration, communication, critical thinking, data analysis, debugging, leadership, modeling, problem solving, project management, and troubleshooting
\end{itemize}

\ProfileSectionHeader[graduation-cap]{bridgingdark}{Experience Requirements}
\fieldlabel{Education:} 
\begin{itemize}
  \item Degrees: Bachelor, Master, or PhD
  \item Disciplines: Applied Science and Technology, Electrical Engineering, or Physics
\end{itemize}

\fieldlabel{On-the-job training:}
\begin{itemize}
  \item Attendance at professional conferences
  \item One-on-one mentoring and collaborative learning with experienced team members
  \item Participation in external courses
\end{itemize}

\fieldlabel{Prior experience:}
\begin{itemize}
  \item Experience at another company or an industry internship is expected
\end{itemize}
\end{tcolorbox}
}

\newcommand{\ProfileBridgingFourteen}{%
\begin{tcolorbox}[
  enhanced, breakable, sharp corners,
  colback=bridgingdark!5, colframe=bridgingdark!70,
  boxrule=0.4pt, arc=2mm,
  borderline west={3pt}{0pt}{bridgingdark},
  left=10pt, right=10pt, top=8pt, bottom=8pt,
  title={\centering\sffamily\large\textcolor{white}{Profile B2.2: Quantum Computer Operators}},
  fonttitle=\sffamily\large,
  coltitle=white,
]
\begin{center}
  {\scriptsize\textcolor{gray!}{\itshape Findings in this profile are not presented in order of importance and will be refined as additional interviews are conducted}}
\end{center}
\vspace{4pt}
{\small
\begin{itemize}[label={}, leftmargin=1.5em, itemsep=1pt, topsep=1pt]
  \item \textbf{Individual positions:} System Operator
  \item \textbf{Company types:} Quantum computing hardware (platform: trapped-ion), Quantum algorithms and software
\end{itemize}
}

\ProfileSectionHeader[building]{bridgingdark}{Occupation-Specific Information}
\fieldlabel{Description:} Quantum computer operators operate, maintain, and schedule software tasks to run on
quantum hardware.

\fieldlabel{Tasks:} On the job, they would...
\begin{itemize}
  \item Deploy, operate, monitor, and maintain quantum systems
  \item Identify and resolve operational issues
\end{itemize}

\ProfileSectionHeader[brain]{bridgingdark}{Worker Requirements: Knowledge, Skills, Abilities (KSAs)}
\fieldlabel{Knowledge of:}
\begin{itemize}
  \item Statistics
\end{itemize}

\fieldlabel{Occupation-specific skills and abilities:}
\begin{itemize}
  \item Able to explain the function and purpose of calibration routines
  \item Able to explain the function and purpose of lasers
  \item Able to identify when operations hit edge cases
  \item Able to interpret imperfect feedback signals
\end{itemize}

\fieldlabel{General skills and abilities:}
\begin{itemize}
  \item Collaboration, communication, critical thinking, data analysis, and problem solving
\end{itemize}

\ProfileSectionHeader[graduation-cap]{bridgingdark}{Experience Requirements}
\fieldlabel{Education:} 
\begin{itemize}
  \item Degree: Bachelor
  \item Disciplines: Engineering or Physics
\end{itemize}

\fieldlabel{On-the-job training:}
\begin{itemize}
  \item Training on operational processes provided on the job
\end{itemize}

\fieldlabel{Prior experience:} 
\begin{itemize}
  \item None required
\end{itemize}
\end{tcolorbox}
}

\subsection*{B1: Bridging Technical Applications \& Software/Hardware}
\addcontentsline{toc}{subsection}{B1: Bridging Technical Applications \& Software/Hardware}

Bridging technical applications with quantum hardware and quantum software involves understanding the applications of quantum technologies and finding ways to implement specific quantum hardware or software in pursuit of that application. 
\subsubsection*{B1.1 Quantum Software Application Developers \& Trainers
}
\addcontentsline{toc}{subsubsection}{B1.1 Quantum Software Application Developers \& Trainers}
\ProfileBridgingEleven
\clearpage

\subsubsection*{B1.2 Quantum Technology End Users
}
\addcontentsline{toc}{subsubsection}{B1.2 Quantum Technology End Users}
\ProfileBridgingSeven
\clearpage

\subsection*{B2: Bridging Hardware \& Software}
\addcontentsline{toc}{subsection}{B2: Bridging Hardware \& Software}

Roles bridging hardware and software use software or computational tools to model quantum systems and make actionable recommendations to hardware specialists or work on the code that controls and runs on quantum hardware.  In either case, these roles require a mix of hardware and software expertise.

\subsubsection*{B2.1 Device \& System Hardware Computational Scientists
}
\addcontentsline{toc}{subsubsection}{B2.1 Device \& System Hardware Computational Scientists}
\ProfileBridgingThirteen
\clearpage
\subsubsection*{B2.2 Quantum Computer Operators
}
\addcontentsline{toc}{subsubsection}{B2.2 Quantum Computer Operators}
\ProfileBridgingFourteen


\clearpage
\section*{Public-Facing and Business Roles Profiles}
\addcontentsline{toc}{section}{Public-Facing and Business Roles Profiles}
The public-facing and business category consists of roles focused on business strategy, leadership, partnerships, public engagement, and government relations.

\newcommand{\ProfilePublicOne}{%
\begin{tcolorbox}[
  enhanced, breakable, sharp corners,
  colback=publicdark!5, colframe=publicdark!70,
  boxrule=0.4pt, arc=2mm,
  borderline west={3pt}{0pt}{publicdark},
  left=10pt, right=10pt, top=8pt, bottom=8pt,
  title={\centering\sffamily\large\textcolor{white}{Profile P1.1: Company Executives}},
  fonttitle=\sffamily\large,
  coltitle=white,
]
\begin{center}
  {\scriptsize\textcolor{gray!}{\itshape Findings in this profile are not presented in order of importance, nor does any individual position in the data include all listed elements.}}
\end{center}
\vspace{4pt}
{\small
\begin{itemize}[label={}, leftmargin=1.5em, itemsep=1pt, topsep=1pt]
  \item \textbf{Individual positions:} Chief Operating Officer, Founder/Scientist at Quantum Startup
  \item \textbf{Company types:} Quantum computing hardware, Quantum networking and communication, Enabling technologies
\end{itemize}
}

\ProfileSectionHeader[building]{publicdark}{Occupation-Specific Information}
\fieldlabel{Description:} Company executives make strategic decisions and are responsible for vision, growth,
and overall management of the organization.

\fieldlabel{Tasks:} On the job, they would...
\begin{itemize}
  \item Create the company roadmap
  \item Develop financial projections
  \item Guide research and product development
  \item Make key decisions for the company
  \item Write proposals
\end{itemize}

\ProfileSectionHeader[brain]{publicdark}{Worker Requirements: Knowledge, Skills, Abilities (KSAs)}
\fieldlabel{Knowledge of:}
\begin{itemize}
  \item Specialized technical knowledge aligned with the company's mission 
  \item Statistics
\end{itemize}

\fieldlabel{Occupation-specific skills and abilities:}
\begin{itemize}
  \item Able to do some programming generally, without specifying whether it is for data processing, data representation, or other purposes
  \item Able to apply specialized knowledge  to guide technical and operational activities in a startup environment
\end{itemize}

\fieldlabel{General skills and abilities:}
\begin{itemize}
  \item Collaboration, communication, critical thinking, data analysis, general experimental skills, leadership, problem solving, project management, and troubleshooting
\end{itemize}

\ProfileSectionHeader[graduation-cap]{publicdark}{Experience Requirements}
\fieldlabel{Education:}
\begin{itemize}
  \item Degrees: Bachelor, Master, or PhD
  \item Disciplines: Chemistry, Computer Science, Engineering, or Physics
\end{itemize}

\fieldlabel{On-the-job training:}
\begin{itemize}
  \item External courses, workshops, and mentoring in business development
  \item One-on-one experiential learning and mentoring
  \item Participation in quantum conferences and workshops
\end{itemize}

\fieldlabel{Prior experience:}
\begin{itemize}
  \item Candidates without a PhD are expected to have relevant work experience
  \item With a PhD, prior industry experience is not required
\end{itemize}
\end{tcolorbox}
}

\newcommand{\ProfilePublicTwo}{%
\begin{tcolorbox}[
  enhanced, breakable, sharp corners,
  colback=publicdark!5, colframe=publicdark!70,
  boxrule=0.4pt, arc=2mm,
  borderline west={3pt}{0pt}{publicdark},
  left=10pt, right=10pt, top=8pt, bottom=8pt,
  title={\centering\sffamily\large\textcolor{white}{Profile 2.1: Hardware Applications \& Technical Sales Specialists}},
  fonttitle=\sffamily\large,
  coltitle=white,
]
\begin{center}
  {\scriptsize\textcolor{gray!}{\itshape Findings in this profile are not presented in order of importance, nor does any individual position in the data include all listed elements.}}
\end{center}
\vspace{4pt}
{\small
\begin{itemize}[label={}, leftmargin=1.5em, itemsep=1pt, topsep=1pt]
  \item \textbf{Individual positions:} Quantum Application Scientist (x2), Application Scientist for Quantum Technologies, Technical Sales (x2), Senior Product Manager, Quantum Systems Engineer
  \item \textbf{Company types:} Quantum computing hardware, Quantum algorithms and software, Quantum sensing, Quantum networking and communication, Enabling technologies
\end{itemize}
}

\ProfileSectionHeader[building]{publicdark}{Occupation-Specific Information}
\fieldlabel{Description:} Hardware applications \& technical sales specialists understand the customer's needs and translate them into solutions using quantum hardware. They facilitate the implementation of their company’s products in the customer’s context (e.g., enabling technologies). They combine technical knowledge with sales and relationship building skills.

\fieldlabel{Tasks:} On the job, they would...
\begin{itemize}
  \item Collaborate with software engineers to improve products and services
  \item Communicate with customers about research projects
  \item Conduct demonstrations at the customer's site
  \item Interface with potential customers and explain product benefits
  \item Provide technical sales support and manage existing products
  \item Propose new product development ideas
  \item Respond to urgent emails and perform administrative work
  \item Stay up to date on advances in quantum information
\end{itemize}

\ProfileSectionHeader[brain]{publicdark}{Worker Requirements: Knowledge, Skills, Abilities (KSAs)}
\fieldlabel{Knowledge of:}
\begin{itemize}
  \item Complex system integration
  \item Cooling and shelving transitions, phase noise, clock architectures, and tradeoffs between atomic or molecular species
  \item Customer application areas (e.g., routing, switching, long-haul quantum communications)
  \item Photonics-based communication, sensing, and computing
  \item Pre-and post processing methods for noisy quantum devices, including error mitigation, error suppression, and dynamic decoupling
  \item Qubit hardware and measurement setups
  \item Quantum gates
  \item Quantum information theory
  \item Statistics
  \item The distinction between classical and quantum behaviors, including entanglement distribution and maintaining quantum state integrity
  \item The quantum technology market and key players
\end{itemize}

\fieldlabel{Occupation-specific skills and abilities:}
\begin{itemize}
  \item Able to adapt to changing requirements and environments
  \item Able to do some programming generally, without specifying whether it is for data processing, data representation, or other purposes
  \item Able to interface with customers and explain technical benefits
  \item Able to provide technical expertise for quantum experiment
  \item Able to read and understand scientific papers
  \item Able to understand customer challenges and map them to technical solutions
\end{itemize}

\fieldlabel{General skills and abilities:}
\begin{itemize}
  \item Collaboration, communication, critical thinking, data analysis, general experimental skills, leadership, organization, problem solving, product management, project management, and troubleshooting
\end{itemize}

\ProfileSectionHeader[graduation-cap]{publicdark}{Experience Requirements}
\fieldlabel{Education:} 
\begin{itemize}
  \item Degrees: Master or PhD
  \item Disciplines: Chemistry, Computer Science, Mathematics, Optics, Physics, or Systems Engineering
\end{itemize} 

\fieldlabel{On-the-job training:}
\begin{itemize}
  \item External sales training, especially for scientists
  \item Management training
  \item One-on-one mentoring and shadowing senior staff
  \item Online lectures, journal reading, workshops, and conference participation
\end{itemize}

\fieldlabel{Prior experience:}
\begin{itemize}
  \item Experimental research experience is highly valued
  \item Internships and postdoctoral experience are valued
  \item Industry and customer-facing roles are preferred but not required
  \item Non-PhD candidates typically need 5+ years relevant experience
  \item PhD may be sufficient without prior work experience
\end{itemize}
\end{tcolorbox}
}

\newcommand{\ProfilePublicThree}{%
\begin{tcolorbox}[
  enhanced, breakable, sharp corners,
  colback=publicdark!5, colframe=publicdark!70,
  boxrule=0.4pt, arc=2mm,
  borderline west={3pt}{0pt}{publicdark},
  left=10pt, right=10pt, top=8pt, bottom=8pt,
  title={\centering\sffamily\large\textcolor{white}{Profile P2.2: Business \& Partnerships Specialists}},
  fonttitle=\sffamily\large,
  coltitle=white,
]
\begin{center}
  {\scriptsize\textcolor{gray!}{\itshape Findings in this profile are not presented in order of importance, nor does any individual position in the data include all listed elements.}}
\end{center}
\vspace{4pt}
{\small
\begin{itemize}[label={}, leftmargin=1.5em, itemsep=1pt, topsep=1pt]
  \item \textbf{Individual positions:} Quantum Business Development Specialist, Quantum Engagement Manager, Business Developer in Quantum Software, Quantum Product Manager
  \item \textbf{Company types:} Quantum computing hardware (platform: superconducting), Quantum algorithms and software, Enabling technologies, Consultants
\end{itemize}
}

\ProfileSectionHeader[building]{publicdark}{Occupation-Specific Information}
\fieldlabel{Description:} Business \& partnerships specialists focus on strategic business development, partnerships, and market expansion in the quantum space.

\fieldlabel{Tasks:} On the job, they would...
\begin{itemize}
  \item Act as liaison between clients, projects, and programs
  \item Conduct business development and create connections
  \item Deliver presentations and public talks
  \item Educate corporations on integrating quantum computing into digital transformation roadmaps
  \item Negotiate business deals and partnerships
  \item Plan commercial activities and marketing events
  \item Translate client use cases into product requirements
  \item Write scientific proposals and white papers for marketing
\end{itemize}

\ProfileSectionHeader[brain]{publicdark}{Worker Requirements: Knowledge, Skills, Abilities (KSAs)}
\fieldlabel{Knowledge of:}
\begin{itemize}
  \item Commercial, legal, and logistical processes
  \item General quantum mechanics and quantum information theory
  \item High-level quantum computing applications
  \item Quantum computation and technology implementation
  \item Statistics
  \item Superconductivity
\end{itemize}

\fieldlabel{Occupation-specific skills and abilities:}
\begin{itemize}
  \item Able to adapt to changing requirements and situations
  \item Able to discuss research scope and depth with partner institutions
  \item Able to do some programming generally, without specifying whether it is for data processing, data representation, or other purposes
  \item Able to identify practical quantum computing applications
  \item Able to map product capabilities to technical benefits
  \item Able to negotiate technically with partners and clients
  \item Able to translate customer technical requirements into specifications
  \item Able to write clearly and effectively for technical and non-technical audiences
\end{itemize}

\fieldlabel{General skills and abilities:}
\begin{itemize}
  \item Collaboration, communication, critical thinking, data analysis, data science, general experimental skills, leadership, problem solving, project management, resource management, storytelling, time management, and writing
\end{itemize}

\ProfileSectionHeader[graduation-cap]{publicdark}{Experience Requirements}
\fieldlabel{Education:} 
\begin{itemize}
  \item Degrees: Bachelor, Master, or PhD 
  \item Disciplines: Chemistry, Computer Science, Engineering, Mathematics, or Physics
\end{itemize}

\fieldlabel{On-the-job training:}
\begin{itemize}
  \item One-on-one learning with engineers about products
  \item Online courses to build quantum-related knowledge
  \item Self-learning via online resources
  \item Workshops to build a well-rounded skill set
\end{itemize}

\fieldlabel{Prior experience:}
\begin{itemize}
  \item Prior work in technical business development, sales, or marketing is common
  \item Some candidates come directly from PhD programs
\end{itemize}
\end{tcolorbox}
}

\newcommand{\ProfilePublicFour}{%
\begin{tcolorbox}[
  enhanced, breakable, sharp corners,
  colback=publicdark!5, colframe=publicdark!70,
  boxrule=0.4pt, arc=2mm,
  borderline west={3pt}{0pt}{publicdark},
  left=10pt, right=10pt, top=8pt, bottom=8pt,
  title={\centering\sffamily\large\textcolor{white}{Profile 1.2: Project Overseers}},
  fonttitle=\sffamily\large,
  coltitle=white,
]
\begin{center}
  {\scriptsize\textcolor{gray!}{\itshape Findings in this profile are not presented in order of importance, nor does any individual position in the data include all listed elements.}}
\end{center}
\vspace{4pt}
{\small
\begin{itemize}[label={}, leftmargin=1.5em, itemsep=1pt, topsep=1pt]
  \item \textbf{Individual positions:} Product Manager, Project Manager
  \item \textbf{Company types:} Quantum computing hardware (platform: trapped-ion), Quantum algorithms and software, Enabling technologies
\end{itemize}
}

\ProfileSectionHeader[building]{publicdark}{Occupation-Specific Information}
\fieldlabel{Description:} Project overseers manage timelines, resources, and deliverables for projects involving quantum technologies.

\fieldlabel{Tasks:} On the job, they would...
\begin{itemize}
  \item Facilitate inter-team communication
  \item Maintain schedules and oversee budgets
  \item Manage product feature roadmaps and service launches
  \item Monitor and manage dependencies between teams
\end{itemize}

\ProfileSectionHeader[brain]{publicdark}{Worker Requirements: Knowledge, Skills, Abilities (KSAs)}
\fieldlabel{Knowledge of:}
\begin{itemize}
  \item Qubit hardware, including capabilities and key performance metrics to evaluate the quality of individuals pieces of hardware and technology 
\end{itemize}

\fieldlabel{Occupation-specific skills and abilities:}
\begin{itemize}
  \item Able to create and maintain documentation for project tracking
  \item Able to support knowledge sharing across teams
\end{itemize}

\fieldlabel{General skills and abilities:}
\begin{itemize}
  \item Communication, critical thinking, problem solving, and project management
\end{itemize}

\ProfileSectionHeader[graduation-cap]{publicdark}{Experience Requirements}
\fieldlabel{Education:} 
\begin{itemize}
  \item Degrees: Bachelor, Master, or PhD
  \item Disciplines: Chemistry, Engineering, Mechanical Engineering, or Physics
\end{itemize} 

\fieldlabel{On-the-job training:}
\begin{itemize}
  \item Occasional online learning opportunities
  \item One-on-one mentoring
\end{itemize}

\fieldlabel{Prior experience:}
\begin{itemize}
  \item Prior project manager experience at another company is common
  \item Transitions from technical roles within the same company are common
\end{itemize}
\end{tcolorbox}
}

\newcommand{\ProfilePublicFive}{%
\begin{tcolorbox}[
  enhanced, breakable, sharp corners,
  colback=publicdark!5, colframe=publicdark!70,
  boxrule=0.4pt, arc=2mm,
  borderline west={3pt}{0pt}{publicdark},
  left=10pt, right=10pt, top=8pt, bottom=8pt,
  title={\centering\sffamily\large\textcolor{white}{Profile P3.1: Education Advocates}},
  fonttitle=\sffamily\large,
  coltitle=white,
]
\begin{center}
  {\scriptsize\textcolor{gray!}{\itshape Findings in this profile are not presented in order of importance, nor does any individual position in the data include all listed elements.}}
\end{center}
\vspace{4pt}
{\small
\begin{itemize}[label={}, leftmargin=1.5em, itemsep=1pt, topsep=1pt]
  \item \textbf{Individual positions:} Content Writer in Quantum, Education Advocate
  \item \textbf{Company types:} Quantum computing hardware (platform: neutral-atom), Quantum algorithms and software, Enabling technologies, Consultants
\end{itemize}
}

\ProfileSectionHeader[building]{publicdark}{Occupation-Specific Information}
\fieldlabel{Description:} Education advocates focus on outreach and education to increase awareness and understanding of quantum technologies.

\fieldlabel{Tasks:} On the job, they would...
\begin{itemize}
  \item Communicate about hardware with external stakeholders
  \item Create news pieces and online journal content
  \item Develop short blog posts and contributed articles
  \item Post and manage content on LinkedIn
  \item Prepare case studies and sales enablement materials
  \item Write informative and accurate quantum content
\end{itemize}

\ProfileSectionHeader[brain]{publicdark}{Worker Requirements: Knowledge, Skills, Abilities (KSAs)}
\fieldlabel{Knowledge of:}
\begin{itemize}
  \item Quantum information science topics such as design and control, quantum gates, qubit hardware, error correction, and quantum cryptography
  \item Quantum software platforms and programming
\end{itemize}

\fieldlabel{Occupation-specific skills and abilities:}
\begin{itemize}
  \item Able to communicate complex quantum concepts clearly and accessibly to a variety of audiences
  \item Able to do some programming generally, without specifying whether it is for data processing, data representation, or other purposes
\end{itemize}

\fieldlabel{General skills and abilities:}
\begin{itemize}
  \item Communication
\end{itemize}

\ProfileSectionHeader[graduation-cap]{publicdark}{Experience Requirements}
\fieldlabel{Education:}
\begin{itemize}
  \item Degree: Bachelor, Master, or PhD
  \item Disciplines: Computer Science, Mathematics, or Physics
\end{itemize}

\fieldlabel{On-the-job training:}
\begin{itemize}
  \item No formal training, but feedback on writing quality is provided with opportunities to revise
\end{itemize}

\fieldlabel{Prior experience:}
\begin{itemize}
  \item About 5 years experience required with a bachelor’s degree
  \item About 2 years experience required with a master’s degree
  \item No prior experience required with a PhD
\end{itemize}
\end{tcolorbox}
}

\newcommand{\ProfilePublicSix}{%
\begin{tcolorbox}[
  enhanced, breakable, sharp corners,
  colback=publicdark!5, colframe=publicdark!70,
  boxrule=0.4pt, arc=2mm,
  borderline west={3pt}{0pt}{publicdark},
  left=10pt, right=10pt, top=8pt, bottom=8pt,
  title={\centering\sffamily\large\textcolor{white}{Profile P3.2: Government-Industry Advocates}},
  fonttitle=\sffamily\large,
  coltitle=white,
]
\begin{center}
  {\scriptsize\textcolor{gray!}{\itshape Findings in this profile are not presented in order of importance and will be refined as additional interviews are conducted}}
\end{center}
\vspace{4pt}
{\small
\begin{itemize}[label={}, leftmargin=1.5em, itemsep=1pt, topsep=1pt]
  \item \textbf{Individual positions:} Government-Industry Advocate
  \item \textbf{Company types:} Quantum computing hardware (platform: neutral-atom), Quantum algorithms and software
\end{itemize}
}

\ProfileSectionHeader[building]{publicdark}{Occupation-Specific Information}
\fieldlabel{Description:} Government-industry advocates communicate industry need to government agencies. They
synthesize technical expertise provided by scientists in their company to write proposals and manage projects across the company.

\fieldlabel{Tasks:} On the job, they would...
\begin{itemize}
  \item Manage projects to get grants submitted
  \item Research systems engineering practices to frame grant proposals
  \item Speak with vendors to understand how company products can be implemented
  \item Work with local and state representatives on implementing quantum-related legislation
\end{itemize}

\ProfileSectionHeader[brain]{publicdark}{Worker Requirements: Knowledge, Skills, Abilities (KSAs)}
\fieldlabel{Knowledge of:}
\begin{itemize}
  \item How scientists in different disciplines think and approach problems
  \item Optoelectronics industry
  \item Laser performance metrics
  \item Quantum information science relevant to their advocacy efforts 
  \item Systems engineering
\end{itemize}

\fieldlabel{Occupation-specific skills and abilities:}
\begin{itemize}
  \item Able to build relationships with government representatives
  \item Able to communicate what quantum is, how it works, and why it matters to policymakers
  \item Able to proactively learn new technical domains as needed
  \item Able to translate research into program and portfolio restructuring
\end{itemize}

\fieldlabel{General skills and abilities:}
\begin{itemize}
  \item Adaptability, communication, community development, coordination, critical thinking, event planning, leadership, and organization
\end{itemize}

\ProfileSectionHeader[graduation-cap]{publicdark}{Experience Requirements}
\fieldlabel{Education:} 
\begin{itemize}
  \item Degree: Bachelor
  \item Discipline: Applied Physics 
\end{itemize}

\fieldlabel{On-the-job training:}
\begin{itemize}
  \item Certification in science policy
  \item Participation in conferences
\end{itemize}

\fieldlabel{Prior experience:}
\begin{itemize}
  \item Previous industry experience is expected
\end{itemize}
\end{tcolorbox}
}


\subsection*{P1: Leadership}
\addcontentsline{toc}{subsection}{P1: Leadership}
Leadership roles are focused on determining company direction and management of the various projects and initiatives the company undertakes. These leadership roles are typically focused on the broader strategic picture of the business and may not require as much technical expertise as the managers or leads within the Hardware and Software categories.
\subsubsection*{P1.1 Company Executives
}
\addcontentsline{toc}{subsubsection}{P1.1 Company Executives}
\ProfilePublicOne
\clearpage
\subsubsection*{P1.2 Project Overseers}
\addcontentsline{toc}{subsubsection}{P1.2 Project Overseers}
\ProfilePublicFour
\clearpage

\subsection*{P2: Client Interactions}
\addcontentsline{toc}{subsection}{P2: Client Interactions}

The client interactions category includes roles focused on coordinating with groups outside of the company for sales or the implementation of products in clients' contexts.

\subsubsection*{P2.1 Hardware Applications \& Technical Sales Specialists}
\addcontentsline{toc}{subsubsection}{P2.1 Hardware Applications \& Technical Sales Specialists}
\ProfilePublicTwo
\clearpage
\subsubsection*{P2.2 Business \& Partnerships Specialists}
\addcontentsline{toc}{subsubsection}{P2.2 Business \& Partnerships Specialists}
\ProfilePublicThree
\clearpage

\subsection*{P3: Engagement}
\addcontentsline{toc}{subsection}{P3: Engagement}
The engagement category consists of roles focused on interfacing with the public or government entities for the purposes of education and advocacy for and about the QISE industry. 
\subsubsection*{P3.1 Education Advocates}
\addcontentsline{toc}{subsubsection}{P3.1 Education Advocates}
\ProfilePublicFive
\clearpage
\subsubsection*{P3.2 Government-Industry Advocates}
\addcontentsline{toc}{subsubsection}{P3.2 Government-Industry Advocates}
\ProfilePublicSix

\newpage
\section*{Acknowledgments}
\addcontentsline{toc}{section}{Acknowledgments}
Thank you to the quantum industry managers and employees who participated in our interviews. 

This material is based on work supported by the National
Science Foundation under Grant Nos. PHY-2333073 and PHY-2333074.

This material is also based on work supported by the Army Research Office and was accomplished under Award Number: W911NF-24-1-0132. The views and conclusions contained in this document are those of the authors and should not be interpreted as representing the official policies, either expressed or implied, of the Army Research Office or the U.S. Government. The U.S. Government is authorized to reproduce and distribute reprints for Government purposes notwithstanding any copyright notation herein.

\vspace{0.9\baselineskip}
{\sffamily\bfseries\color{brandnavy} About the project}\\
This report is part of a collaborative project between researchers at the University of Colorado Boulder \& Rochester Institute of Technology to advance quantum information science education and strengthen the quantum workforce.  
Learn more about the broader effort at
\href{https://www.rit.edu/quantumeducationandworkforce/}{rit.edu/quantumeducationandworkforce}.

\vspace{0.9\baselineskip}
{\sffamily\bfseries\color{brandnavy} Contact information}\\
For questions about this report or the broader project, please contact the principal investigators:

\noindent\hfill
\begin{minipage}{0.9\linewidth}
\centering
\begin{tabular}{@{}p{0.45\linewidth}p{0.45\linewidth}@{}}
Heather J.~Lewandowski & Benjamin M.~Zwickl \\
University of Colorado Boulder & Rochester Institute of Technology \\
\href{mailto:lewandoh@colorado.edu}{lewandoh@colorado.edu} &
\href{mailto:bmzsps@rit.edu}{bmzsps@rit.edu} \\
\end{tabular}
\end{minipage}
\hfill\mbox{}

\vspace{0.9\baselineskip}
{\sffamily\bfseries\color{brandnavy} Suggested citation}\\
Shams El-Adawy, A.\,R. Pi\~na, Benjamin M.\,Zwickl, \& H. J. Lewandowski (January 2026).
\emph{Profiles of Roles in the Quantum Industry}
(Quantum Workforce Report Series, Report 3). University of Colorado Boulder \& Rochester Institute of Technology.

\newpage
\printbibliography
\end{document}